\documentclass[prd,showpacs,preprintnumbers,nofootinbib,eqsecnum]{revtex4}

\usepackage[dvipdfm]{graphicx}
\usepackage{dcolumn}
\usepackage{bm}
\usepackage{latexsym}
\usepackage{amsfonts}
\usepackage{amssymb}
\usepackage{amsmath}

\begin{document}

\preprint{YITP-11-67, KUNS-2348}

\title{
Holographic Dual of de Sitter Universe with AdS Bubbles 
}
\author{Sugumi Kanno$^{1}$}
\author{Misao Sasaki$^{2,4}$}
\author{Jiro Soda$^{3}$}
\affiliation{$^{1}$Institute of Cosmology, Department of Physics and Astronomy, 
Tufts University, Medford, Massachusetts 02155, USA \\
 $^{2}$Yukawa Institute,  Kyoto University, Kyoto 606-8502, Japan \\
 $^{3}$Department of Physics,  Kyoto University, Kyoto 606-8502, Japan\\
$^{4}$Korea Institute for Advanced Study,
Seoul 130-722, Republic of Korea
}%

\date{\today}

\begin{abstract}
We study the proposal that a de Sitter (dS) universe with an 
Anti-de Sitter (AdS) bubble
 can be replaced by a dS universe with a boundary CFT. 
To explore this duality, we consider incident gravitons coming 
from the dS universe through the bubble wall into the AdS bubble 
in the original picture. 
In the dual picture, this process
has to be identified with the absorption of gravitons by CFT 
matter. We have obtained a general formula for the absorption 
probability in general $d+1$ spacetime dimensions. The result shows
the different behavior depending on whether spacetime dimensions
are even or odd.
We find that the absorption process of gravitons from
the dS universe by CFT matter is controlled by localized
gravitons (massive bound state modes in the Kaluza-Klein decomposition)
in the dS universe.
The absorption probability is determined by 
the effective degrees of freedom of the CFT matter
and the effective gravitational coupling constant
 which encodes information of localized gravitons. 
We speculate that the dual of $(d+1)$-dimensional dS universe with 
an AdS bubble is also dual to a $d$-dimensional dS universe with 
CFT matter.
\end{abstract}

\maketitle
\section{Introduction}

The resolution of the spacetime singularity has been a long standing
problem~\cite{Craps:2010bg}. It is believed that quantum gravity 
will resolve the singularity, and string theory seems the most promising 
candidate for quantum gravity.
So far, one of the most important outcomes from string theory is the AdS/CFT
correspondence~\cite{Maldacena:1997re, Gubser:1998bc, Witten:1998qj},
or more generally the Gauge/Gravity correspondence. 

There are two ways to use the AdS/CFT correspondence.
The first one is to use the correspondence to understand
the strongly coupled field theory using classical gravity. Remarkably, 
the correspondence has been used in the condensed matter context
for calculating correlation functions in a strongly 
interacting field theory using the dual classical gravity description. 
The AdS/CFT correspondence can also be used the other way around.
Namely, the AdS/CFT correspondence can be used to describe
quantum gravity using a well-defined dual field theory.
Since CFT is a healthy, unitary theory, it is natural to expect 
that the AdS/CFT correspondence resolves the singularity problem.
Indeed, this view has been applied to the singularity problem in the 
past~\cite{Hertog:2005hu,Hertog:2004rz,Craps:2007ch,Craps:2009qc}.

Even in the classical context, the singularity is annoying.
Recently, it is found that black hole singularities can be treated 
by an excision method in numerical relativity~\cite{Pretorius:2005gq}. 
It is interesting to see if this idea can be promoted to a theoretical
framework.
In fact, the singularity excision has been discussed in the context of 
black hole evaporation~\cite{Horowitz:2005vp,Murata:2007jh}.
In principle, the excision method can be formulated as an effective 
field theory with boundary,
which can be improved with a renormalization group idea. 

Interestingly, the above apparently two different approaches merged 
into a recent proposal by Maldacena that an AdS bubble 
in a dS spacetime can be replaced by a boundary CFT according 
to the AdS/CFT duality~\cite{Maldacena:2010un}. 
Since AdS spacetime is unstable to the formation of a singularity
under a small perturbation, if this proposal is valid, we can avoid
dealing directly with a singular spacetime by replacing it
with the spacetime with boundary.
In a sense, we have a stringy realization of the singularity excision. 

Apparently, it is worth exploring this proposal in detail.
A simple, direct route to the understanding of the proposal is to study
incident gravitons from the outside dS universe into the AdS bubble.
In the original picture, gravitons can penetrate into the AdS bubble.
One can calculate the transmission probability for this process.
How should we understand this process in the dual picture in which
the AdS bubble is replaced by the boundary bubble wall with the dual CFT?
Since there is nothing inside of the bubble wall, 
it is expected intuitively that the energy of gravitons would be
transferred to particles of the CFT. 

In fact in the limit of infinitely large dS radius (i.e., in the Minkowski
limit), by calculating the transmission probability of gravitons 
coming from the dS side through the bubble wall, it is argued that 
the result can be interpreted as the absorption probability of gravitons 
into CFT particles~\cite{Garriga:2010fu}. However, the method used 
in~\cite{Garriga:2010fu} was specific to four dimensions and the detailed
discussion was provided only for the case of an AdS bubble in Minkowski 
spacetime. 

Here we are interested in more general situations.
If Maldacena's conjecture is correct, it should hold in arbitrary dimensions.
We study the singularity excision in general $d+1$ spacetime dimensions.
In particular, we give much attention to the role of the dS universe
in this dual picture.
We note that, in the string landscape or in the multiverse
picture~\cite{Susskind:2003kw,Freivogel:2006xu,Garriga:2008ks},
AdS bubbles will be formed in various kinds of dS vacua in (perhaps) various
spacetime dimensions. Our analysis will be relevant to 
these cosmologically interesting situations.

The organization of this paper is as follows.
In section~\ref{sec:model}, we provide a model for AdS bubbles 
nucleating in the inflating multiverse and explain the dual picture 
where the AdS bubble is replaced with the boundary CFT. 
In section~\ref{sec:transprob}, we calculate the transmission probability
in arbitrary dimensions. In section~\ref{sec:dual},
we identified the transmission probability with the 
absorption probability in the dual picture.
In the decoupling limit of gravity in the sense of AdS/CFT correspondence, 
it turns out that the absorption probability is factorized into 
a factor from effective degrees of CFT matter,
an effective gravitational coupling constant, and a kinematical factor. 
We obtain an explicit formula for various situations. 
We find the absorption probability is controlled by graviton bound states
originating in de Sitter spacetime. 
The final section is devoted to the conclusion.
In Appendix~\ref{app:Legendre}, we summarize several useful
mathematical formulas.
In Appendix~\ref{app:4Dexact}, we give an exact formula 
for the transmission probability in four dimensions without 
taking the decoupling limit,
in order to make precise comparison with the result 
obtained in~\cite{Garriga:2010fu}.
In Appendix~\ref{app:BS}, a computation of the bound states in 
the presence of an AdS bubble is given.
In Appendix~\ref{app:wallfluc}, we discuss the wall fluctuation mode 
and compare it with bound states.

\section{Holographic excision of AdS bubbles }
\label{sec:model}

In the landscape picture, AdS bubbles nucleate with a 
certain probability in the inflating multiverse~\cite{Coleman:1980aw}.
In this section, we pick up a part of the multiverse where the universe 
is in a dS vacuum. 
First, we describe an AdS bubble in the dS universe. Then, we present
a dual of this geometry by following Maldacena~\cite{Maldacena:2010un}.

\begin{figure}[htbp]
 \begin{center}
  \includegraphics[width=100mm]{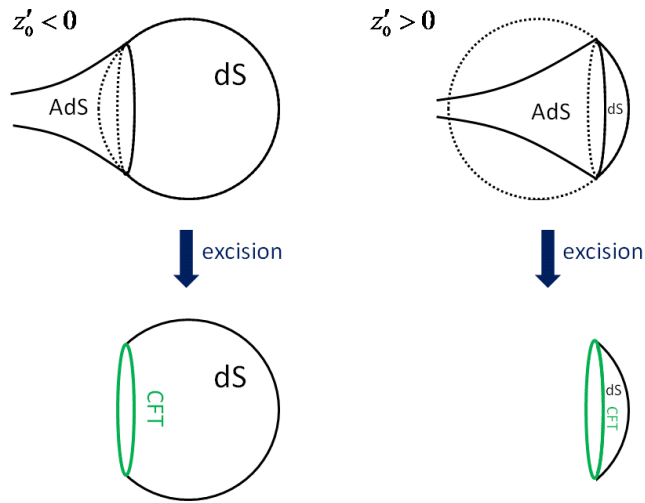}
 \end{center}
 \caption{The excision process is depicted. In the case $z_0'<0$,
 the AdS bubble does not hide the great circle of dS.
 While, in the case $z_0'>0$, there is no great circle in dS 
spacetime. The former describes an AdS bubble in the dS universe. 
The latter describes the case when both AdS and dS spacetimes
are surrounded by a common boundary wall.}
 \label{fig:1}
\end{figure}

We begin with the background spacetime 
with a $(d+1)$-dimensional AdS bubble of curvature radius $\ell_{\rm AdS}$ 
in a dS spacetime of curvature radius $\ell_{\rm dS}$.
The metric we consider is
\begin{eqnarray}
ds^2 = a^2 (z) \left[ dz^2 +  ds_{\rm dS}^2 \right]  \ ,
\label{metric3}
\end{eqnarray}
where $ds_{\rm dS}^2$ is the $d$-dimensional dS spacetime
with unit curvature radius. Inside the bubble, the conformal factor 
is given by the AdS spacetime,
\begin{eqnarray}
a(z) = \frac{\ell_{\rm AdS}}{\sinh (z_0-z)}
\,,\hspace{1cm}
z\leq 0\,.
\label{metric:ads}
\end{eqnarray}
This region is separated by a domain wall from an outer region 
of the dS universe. Then outside the bubble is expressed by the dS
spacetime,
\begin{eqnarray}
a(z) = \frac{\ell_{\rm dS}}{\cosh (z+z_0^\prime)}
\,,\hspace{1cm}
z\geq 0\,.
\label{metric:ds}
\end{eqnarray}
Here, $z_0$ and $z_0^\prime$ are the parameters that
determine the AdS and dS curvature radii and the position
of the bubble wall in each spacetime.
We match these two spacetimes at $z=0$.  Note that
$z_0>0$ while $z_0'$ can be either positive or negative.
In Fig.~\ref{fig:1}, we explain the geometrical meaning of the
parameter $z_0'$. 
The continuity of these metrics leads to the relation,
\begin{eqnarray}
\frac{\ell_{\rm AdS} }{\sinh z_0}
 = \frac{\ell_{\rm dS} }{\cosh z'_0} =a_w \,,
\label{continuity1}
\end{eqnarray}
where $a_w$ is the curvature radius of the domain wall
which is a $d$-dimensional dS spacetime.
The continuity condition Eq.~(\ref{continuity1})
gives  the relations  $\sinh z_0 = \ell_{\rm AdS} /a_w$ 
and $\cosh z'_0 = \ell_{\rm dS} /a_w$.
Thus, varying $z_0$ and $z_0'$ is equivalent to varying
$\ell_{\rm AdS}$ and $\ell_{\rm dS}$ under fixed $a_w$. 
The junction condition at the domain wall gives the tension
of the domain wall of the form,
\begin{eqnarray}
T = T_{\rm cr} \frac{\ell_{\rm AdS}}{a_w}\left[ 
\sqrt{1+\left( \frac{a_w}{\ell_{\rm AdS}}\right)^2 }
\pm
\sqrt{1-\left( \frac{a_w}{\ell_{\rm dS}}\right)^2 } 
\right]\,,
\hspace{1cm}
T_{\rm cr} = \frac{d-1}{\kappa^2~\ell_{\rm AdS}}\,,
\label{tension}
\end{eqnarray}
where the plus and minus signs in front of the second term
correspond to the choice $z_0'>0$ and $z_0'<0$, respectively. 
We have defined $\kappa^2=8\pi G_{d+1} $ and $T_{\rm cr}$ is the tension
 corresponding to a flat domain wall $a_w \rightarrow \infty$.
We note that the choice $z^\prime_0<0$ is more natural in the sense that
it corresponds to the Coleman-De Luccia bubble~\cite{Coleman:1980aw}, 
that is, an AdS bubble nucleated in the dS universe. 
On the other hand, the choice $z_0'>0$ corresponds to the case 
when both AdS and dS spacetimes are surrounded by a common boundary wall. 
Although the latter may not be easily realizable 
if considered in the context of field theory,
mathematically there is nothing wrong with such a configuration.
Moreover, if we consider multiple nucleation processes,
a dS bubble could be produced as a remnant of the false 
vacuum~\cite{Sato:1981bf}.
Therefore we consider both of these two cases in the following.

By following Maldacena's conjecture~\cite{Maldacena:2010un}, 
we excise the AdS bubble and replace it with a boundary CFT.
The resultant spacetime is dS bounded
by a boundary dS on which CFT lives. 
There exists no spacetime beyond the bubble wall (see Fig.~\ref{fig:1}).
In this paper, we study consequences of this conjectured duality 
through probe gravitons. 

In order to distinguish the outer dS universe from the boundary dS 
where CFT resides, we refer the outer dS universe as dS bulk in 
the following.

\section{Transmission Probability in Arbitrary Dimensions}
\label{sec:transprob}

In this section, we calculate the transmission probability of incident 
gravitons from the dS bulk into the AdS bubble in arbitrary 
dimensions. 

We consider gravitational waves on the background explained in the
previous section.
The conformal factor in this background is expressed by using
the theta function as
\begin{eqnarray}
a(z)=\theta(-z)~\frac{\ell_{\rm AdS}}{\sinh (z_0-z)}
+\theta(z)~\frac{\ell_{\rm dS}}{\cosh (z+z_0^\prime)}
\,.
\end{eqnarray}
The $d$-dimensional spacetime components 
$(i,j)$ of the tensor perturbation, $\psi_{ij}$, are defined by
\begin{eqnarray}
ds^2 = a^2 (z) \left[ dz^2 +  ds_{\rm dS}^2 + \psi_{ij} dx^i dx^j \right]\,.
\label{metric+psi}
\end{eqnarray}
The wave equation for $\psi_{ij}$  is given by
\begin{eqnarray}
\left[
\frac{d^2}{dz^2} + \frac{d-1}{a}\frac{da}{dz}\frac{d}{dz}
+\left(\Box - 2\right)
\right]\psi_{ij} = 0\,.
\end{eqnarray}
Here, $\Box$ is the $d$-dimensional d'Alembertian operator.
After separation
of variables by $\psi_{ij}(z,x^i) = \tilde{\psi}(z) Y_{ij}(x^i) $,
we find $Y_{ij}$ satisfies
\begin{eqnarray}
\left[ \Box -2 -M^2 \right] Y_{ij} =0  \, , 
\quad\quad \nabla^jY_{ij}=Y^i{}_i =0 \, ,
\label{TTmode}
\end{eqnarray}
where $M$ is a constant of separation which represents the
Kaluza-Klein (KK) mass of the gravitons in $d$ dimensions.
Changing the variable: $\tilde{\psi}=\phi/a^{(d-1)/2}$, we obtain a 
Schr\"{o}dinger type wave equation,
\begin{eqnarray}
\left[-\frac{d^2}{dz^2} + V(z)\right] \phi = M^2 \phi\,,
\label{basic}
\end{eqnarray}
where the effective potential $V$ reads
\begin{eqnarray}
V(z) &=&\frac{1}{a^{(d-1)/2}}\frac{d^2a^{(d-1)/2}}{dz^2}
=\frac{d-1}{2a}\frac{d^2a}{dz^2}
+\frac{(d-1)(d-3)}{4a^2}\left(\frac{da}{dz}\right)^2
\nonumber\\
&=&
-\frac{d-1}{2}
\Bigl[\coth z_0 + \tanh z^\prime_0\Bigr]\delta (z)
\nonumber\\
&&+\theta (-z) \left[\,\frac{(d-1)^2}{4}
+\frac{d^2-1}{4\sinh^2(z-z_0)}
\,\right]
+\theta (z) \left[\,
\frac{(d-1)^2}{4}
-\frac{d^2-1}{4\cosh^2(z+z_0^\prime)}\,\right]\ .
\end{eqnarray}
\begin{figure}[ht]
\begin{center}
\begin{minipage}{8.5cm}
\begin{center}
\hspace{-1.5cm}
\includegraphics[width=95mm]{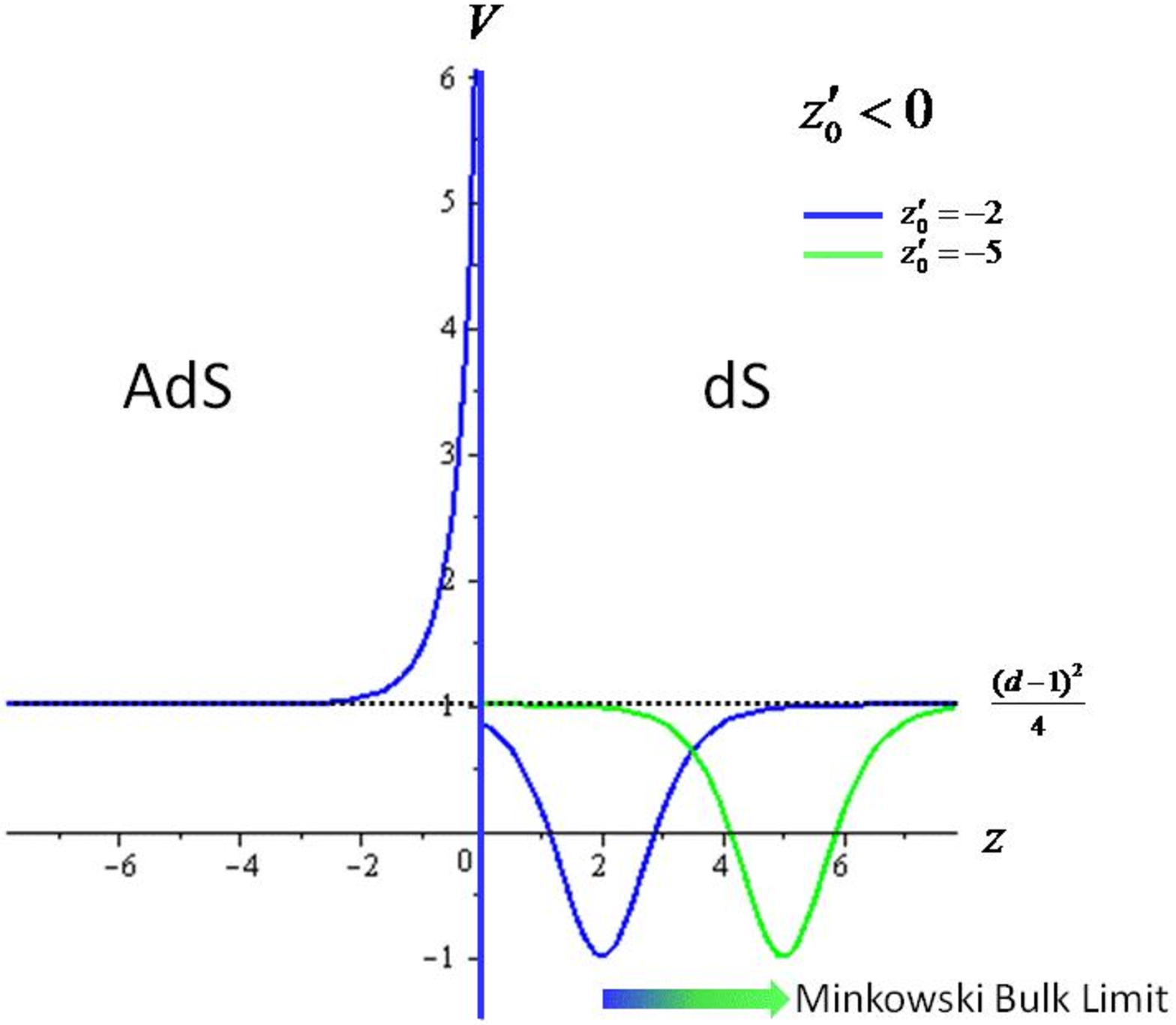}\vspace{0cm}
\caption{The potential for $z^\prime_0<0$.}
\label{fig:2}
\end{center}
\end{minipage}
\hspace{1mm}
\begin{minipage}{8.5cm}
\begin{center}
\hspace{-1.5cm}
\includegraphics[width=95mm]{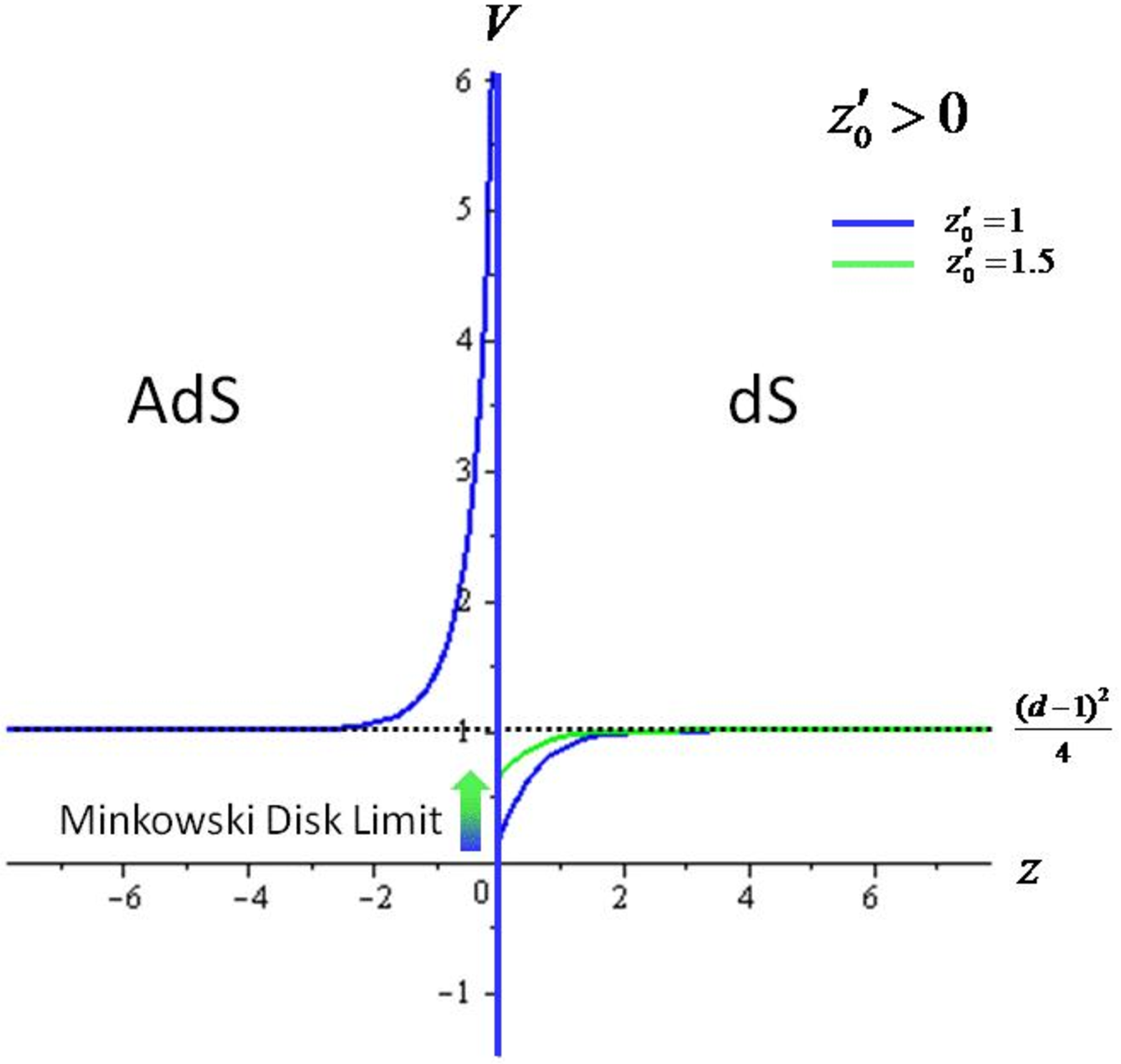}\vspace{0cm}
\caption{The potential for $z^\prime_0>0$.}
\label{fig:3}
\end{center}
\end{minipage}
\end{center}
\end{figure}
Note that the first term proportional to the delta function
corresponds to the tension of the domain wall, Eq.~(\ref{tension}).
In Figs.~\ref{fig:2} and \ref{fig:3}, we depict
the above potential for $z_0' <0$ and $z_0' >0$, respectively.
At the position of the domain wall $z=0$, there is 
an infinitely deep well due to a minus of the delta function. 
We can see that, for $z_0'<0$, the potential well in the dS bulk
 is moving away from the position of the domain wall as the dS
radius gets larger and eventually approaches the Minkowski bulk.
On the other hand for $z_0'>0$,
the potential well at the domain wall becomes shallow as 
the dS radius gets larger and eventually approaches
a Minkowski disk.

Now we want to consider incident gravitational waves from the
 dS bulk. The boundary condition we consider is
\begin{eqnarray}
\phi(z) &\to&  {\bm T}\,e^{-ikz}\,\,;
\hspace{1.7cm} z\to-\infty\,,  
\label{bc:ads}\\
\phi(z) &\to& {\bm R}\,e^{ikz} + e^{-ikz}\,\,;
\quad\quad z\to\infty\,,
\label{bc:ds}
\end{eqnarray}
where we have defined 
\begin{eqnarray}
k\equiv\sqrt{M^2-\frac{(d-1)^2}{4}}
\quad\leftrightarrow\quad
M^2=k^2+\frac{(d-1)^2}{4}\,,
\label{defk}
\end{eqnarray}
and $\bm{T}$ and $\bm{R}$ represent the amplitudes of
transmission and reflection waves, respectively.

\subsection{Solutions inside AdS bubble}
\label{subsec:adsbubble}

It is easy to see that the general solution is
given in terms of the associated Legendre functions inside AdS bubble. 
By imposing the boundary
condition~(\ref{bc:ads}), we find the solution is given by
\begin{eqnarray}
\phi(z) &=&c_1~P_{\nu}^{ik}\left(\coth(z_0-z)\right)
\,;\quad\quad c_1\equiv  {\bm T}\,\Gamma(1-ik)\,e^{-ikz_0}\,,\quad
\nu\equiv \frac{d-1}{2}\,.
\label{soln:ads}
\end{eqnarray}
Note that we consider the case $M^2>(d-1)^2/4$,
thus we take an imaginary number $ik$ as the order of the associated Legendre 
function. Since there are several different normalizations
of $P_\nu^\mu(u)$, in order to avoid confusion, we spell out
the definitions adopted in this paper and summarize
other useful formulas related to the associated Legendre functions
in Appendix~\ref{app:Legendre}.

\subsection{Solutions in de Sitter bulk}
\label{subsec:dsbulk}

In the dS bulk, the general solution is also given by the
associated Legendre functions. 
After imposing the boundary condition,~(\ref{bc:ds}), 
we obtain the solution of the form,
\begin{eqnarray}
\phi(z) &=& d_1~P_{\nu}^{ik}\left(\tanh(z+z_0^\prime)\right)
+d_2~P_{\nu}^{-ik}\left(\tanh(z+z_0^\prime)\right)
\,,
\label{soln:ds}
\end{eqnarray}
where $\nu=(d-1)/2$ as before, and
\begin{eqnarray}
d_1\equiv  {\bm R} \,\Gamma(1-ik)\,e^{\pi k}e^{-ikz_0'}
\,,\hspace{1cm}
d_2\equiv \Gamma(1+ik)\,e^{-\pi k}e^{ikz_0'}\,.
\end{eqnarray}

\subsection{Transmission probability}
\label{subsec:transprob}

In order to solve the scattering problem, we first match
the solutions obtained in Eqs.(\ref{soln:ads}) and (\ref{soln:ds}) 
smoothly at the domain wall.
This requires the following two junction conditions at $z=0$: 
\begin{eqnarray}
&&c_1~P^{ik}_{\nu}\left(x_0\right)
= d_1~P^{ik}_{\nu}\left(x_0^\prime\right)
+d_2~P^{-ik}_{\nu}\left(x_0^\prime\right)
\label{smooth1}
\,,\\
&&c_1~\left[
\left(\nu + 1\right)x_0~P^{ik}_{\nu}(x_0)
-\left(\nu+1-ik\right)P^{ik}_{\nu+1}(x_0)
+\nu\left(x_0+x_0^\prime\right)P^{ik}_{\nu}(x_0)
\right]
\nonumber\\
&&=-d_1\left[
(\nu+1)x_0^\prime~P^{ik}_{\nu}(x_0^\prime)
-\left(\nu+1-ik\right)P^{ik}_{\nu+1}(x_0^\prime)
\right]
-d_2\left[
(\nu+1)x_0^\prime~P^{-ik}_{\nu}(x_0^\prime)
-\left(\nu+1+ik\right)P^{-ik}_{\nu+1}(x_0^\prime)
\right]
\label{smooth2}\,,
\end{eqnarray}
where we have defined 
\begin{eqnarray}
x_0\equiv\coth z_0\,,
\quad\quad
x_0^\prime\equiv\tanh z_0^\prime\,.
\label{x0x0pdef}
\end{eqnarray}
We note that the ranges of these variables are 
$|x_0| > 1$ and $|x_0^\prime|< 1$ by definition.

From the asymptotic forms of Eqs.~(\ref{bc:ads}) and (\ref{bc:ds}), 
we see the transmission probability is given by $|{\bm T}|^2$. 
By solving above continuous
conditions, we find the transmission probability 
\begin{eqnarray}
{\cal T} \equiv |{\bm T}|^2
=\bigg|\frac{c_1}{\Gamma(1-ik)}\bigg|^2
=\frac{4}{\pi^2}\sinh^2 \pi k ~e^{-2\pi k}|{\cal M}|^2
\,,
\label{transmission}
\end{eqnarray}
where
\begin{eqnarray}
{\cal M}^{-1}(x_0, x_0^\prime, k)
&=&P^{ik}_{\nu}(x_0^\prime)\left[
\left(\nu +1 -ik\right)P^{ik}_{\nu+1}(x_0)
-\left(\nu+1\right)x_0~P^{ik}_{\nu}(x_0)
-\nu\left(x_0+x_0^\prime\right)P^{ik}_{\nu}(x_0)
\right]
\nonumber\\
&&+P^{ik}_{\nu}(x_0)\left[
\left(\nu+1-ik\right)P^{ik}_{\nu+1}(x_0^\prime)
-\left(\nu+1\right)x_0^\prime~P^{ik}_{\nu}(x_0^\prime)
\right]
\nonumber\\
&=&-(\nu+ik)\left(P^{ik}_{\nu}(x_0) P^{ik}_{\nu-1}(x_0')
+ P^{ik}_{\nu-1}(x_0)P^{ik}_{\nu}(x_0')\right)\,,
\label{Minv}
\end{eqnarray}
where we have used a recurrence relation: 
$(\nu+1-ik)P^{ik}_{\nu+1}(x)
-(2\nu+1)xP^{ik}_{\nu}(x)
+(\nu+\mu)P^{ik}_{\nu-1}(x)=0$ at the last step.
It should be stressed that this is quite a general result.

\section{Dual Interpretation}
\label{sec:dual}

There are three relevant length scales in the system, namely, 
$\ell_{\rm dS}$,  $\ell_{\rm AdS}$, and $a_w$. Since we know
$\ell_{\rm dS}=a_w\cosh z_0^\prime$
from the junction condition~(\ref{continuity1}), 
we see $\ell_{\rm dS}$ is always larger than $a_w$.
So far we have imposed no restriction on the model parameters.

We are interested in the regime where
physical momenta of gravitons are much smaller than the AdS scale,
$ k/a_w \ll 1/\ell_{\rm AdS}$. This is guaranteed
for a finite value of $k$ if we consider the case of the 
large bubble limit,
\begin{eqnarray}
\frac{a_w}{\ell_{\rm AdS}}=\frac{1}{\sinh z_0} \gg 1 \, ,
\label{assumption}
\end{eqnarray}
which is equivalent to $z_0 \ll 1$
or $x_0 \gg 1$ according to the definition of $x_0$
in Eq.~(\ref{x0x0pdef}). 
This means the case when the dS boundary is located almost
at infinity in the AdS spacetime and hence the decoupling limit 
of gravity from the boundary CFT~\cite{Koyama:2001rf,Li:2011bt}.
There, the AdS/CFT correspondence is supposed to be exact
and we may consider the boundary CFT instead of the gravity 
in the AdS bubble.
This is why we can replace the AdS bubble by the boundary CFT.
Then we can move on to the dual picture of no AdS spacetime inside
the bubble and identify the absorption probability of CFT 
particles ${\cal A}(k)$ with the 
transmission probability of gravitons from the dS bulk ${\cal T}(k)$,
\begin{eqnarray}
    {\cal A} (k) = {\cal T}(k) \, .
\end{eqnarray}

Under the above assumption, Eq.~(\ref{Minv}) is simplified as
\begin{eqnarray}
{\cal M}^{-1}(x_0, x_0^\prime, k)
=  -\left(\nu+ik\right)
P^{ik}_{\nu}(x_0) P^{ik}_{\nu-1}(x_0^\prime)
\,.
\label{M1}
\end{eqnarray}
We notice that the dependences of the AdS bubble 
on $x_0$ and the dS bulk on $x_0'$ are factorized in this case.
Using the asymptotic expansion of $P^{ik}_{\nu}(x_0)$ at $x_0\gg 1$, 
we obtain 
\begin{eqnarray}
|{\cal M}|^{-2}
&=&(\nu^2+k^2)\frac{\sinh^2\pi k}{\pi^2k^2}e^{-2\pi k}
\bigl|F_{\nu,k}(x_0)\bigr|^{2}
\bigl|F_{\nu-1,k}(x_0')\bigr|^{2}
\cr
&\mathop{=}\limits_{x_0\gg1}&
\left(\frac{a_w}{\ell_{\rm AdS}}\right)^{2\nu}
2^{2\nu}\frac{\sinh^2\pi k}{\pi^3}e^{-2\pi k}\,
\frac{\Gamma\left(\nu+\frac{1}{2}\right)}{\Gamma\left(\nu\right)}
\left|\frac{\Gamma(\nu)\Gamma(-ik)}{\Gamma(\nu-ik)}\right|^2
\bigl|F_{\nu-1,k}(x_0')\bigr|^{2}
\,,
\label{Minv2}
\end{eqnarray}
where for notational simplicity 
we have introduced a function $F_{\nu,k}(x)$ by
\begin{eqnarray}
F_{\nu,k}(x)\equiv
F\left(-\nu,~\nu+1,~1-ik;~\frac{1-x}{2}\right)\,,
\label{defF}
\end{eqnarray}
where $F(\alpha,\beta,\gamma,u)$ is a hypergeometric function.
Plugging the above into Eq.~(\ref{transmission}),
the absorption probability is expressed as
\begin{eqnarray}
{\cal A} (k) 
&=&2^{2-2\nu}\pi\left(\frac{\ell_{\rm AdS}}{a_w}\right)^{2\nu}
\frac{\Gamma^2\left(\nu\right)}{\Gamma^2\left(\nu+\frac{1}{2}\right)}
\left|\frac{\Gamma(\nu-ik)}{\Gamma(\nu)\Gamma(-ik)}\right|^2
\bigl|F_{\nu-1,k}(x_0')\bigr|^{-2}
\cr
\cr
&=&2^{2-2\nu}k\sinh\pi k\,
\left(\frac{\ell_{\rm AdS}}{a_w}\right)^{2\nu}
\frac{\Gamma^2\left(\nu\right)}{\Gamma^2\left(\nu+\frac{1}{2}\right)}
\left[ \prod\limits^{\infty}_{n=0} 
\left( 1+\frac{k^2}{\left( \nu +n \right)^2 }\right) \right]^{-1}
\bigl|F_{\nu-1,k}(x_0')\bigr|^{-2}\,.
\label{Aprob}
\end{eqnarray}

Let us consider the case of four dimensions ($d=3$ or $\nu=1$). 
Since the factor $F_{0,k}(x_0')=1$ is independent of $x^\prime_0$,
using the first line of Eq.~(\ref{Aprob}),
the absorption probability is easily found to be
\begin{eqnarray}
{\cal A}=4k^2\left(\frac{\ell_{\rm AdS}}{a_w}\right)^2\,.
\label{A:4D}
\end{eqnarray}
The energy of KK gravitons on the wall
is given by Eq.~(\ref{defk}) with $d=3$,
\begin{eqnarray}
E^2 =\frac{M^2}{a_w^2}= k_w^2 + \lambda_{\rm IR}^2    \,,
\hspace{1cm}
\left(\Delta E\right)^2 = E^2 - \lambda_{\rm IR}^2 = k_w^2
\,,
\end{eqnarray}
where we have defined $k_w = k/a_w$ and $\lambda_{\rm IR}=1/a_w$.
Then Eq.~(\ref{A:4D}) reads
\begin{eqnarray}
{\cal A}=c_4 \ G_4 \left(\Delta E\right)^2
\,,\hspace{1cm}
c_4 =4\frac{\ell^2_{\rm AdS}}{G_4}\,,
\label{A:4DCFT}
\end{eqnarray}
where we have introduced the four-dimensional gravitational 
coupling $G_4$ and degrees of freedom $c_4$ of the CFT according 
to the AdS/CFT 
correspondence~\cite{Maldacena:1997re, Gubser:1998bc, Witten:1998qj}. 
The effect of the AdS spacetime comes in only through this $c_4$.
The factor $G_4 (\Delta E)^2$ is the decay rate of gravitons
into a single degree of freedom of the CFT. 
This result tells us that gravitons coming from the four-dimensional 
dS bulk decay into CFT particles of three dimensions.
This is a standard decay rate in the Minkowski spacetime, which 
is in agreement with the low energy limit of the result obtained
in~\cite{Garriga:2010fu}.\footnote{Note that the result~(\ref{A:4D})
or (\ref{A:4DCFT}) is unbounded as energy increases, 
apparently violating the unitarity bound. However, this is simply
due to our assumption that $k/a_w\ll1/\ell_{\rm AdS}$. If we relax
this assumption, the result is perfectly finite in the limit of
the infinitely large energy.
For completeness, we derive
an exact expression for the transmission probability in four dimensions
without any approximation in Appendix~\ref{app:4Dexact}. It agrees
with \cite{Garriga:2010fu} in the limit of the Minkowski bulk.}

It is known that bound states appear in the transmission
amplitude as poles. In fact, we find the poles in the factor
from the dS bulk $P^{ik}_{\nu-1} (x_0')$ in the complex $k$-plane
in Eq.~(\ref{Aprob}) from the calculation of the bound 
state, Eq.~(\ref{boundstate}), in Appendix~\ref{app:BS}. 
Also curiously, there appear other poles in the factor of the
infinite product coming from the AdS bubble
which are not related to the bound states, depending on
whether spacetime dimensions are even or odd. 
Let us see this next.

In the case of even spacetime dimensions (or odd $d$) equal to
or more than four,  $d=2m+3$ or $\nu=m+1$ ($m=0,1,2,\cdots$), 
Eq.~(\ref{Aprob}) reduces to
\begin{eqnarray}
{\cal A} (k) = 4k^2 \left(\frac{\ell_{\rm AdS}}{a_w}\right)^{2m+2}
\left[ \frac{m!}{(2m+1)!!}\right]^2  
\prod\limits^{m}_{n=1} 
\left( 1+\frac{k^2}{n^2 }\right)
\bigg| F_{m,k}(x_0')\bigg|^{-2} \,,
\label{AB:d=2m+3}
\end{eqnarray}
where we have used the second line of Eq.~(\ref{Aprob})
and rewritten the infinite product in it as
\begin{eqnarray}
\left[ \prod\limits^{\infty}_{n=1}
\left(1+\frac{k^2}{\left(m+n\right)^2}
\right) \right]^{-1}
= \prod\limits^{m}_{n=1} 
\left(1+\frac{k^2}{n^2}
\right)
\frac{\pi k}{\sinh\pi k}\,.
\label{IP:even}
\end{eqnarray}

We see that written in this way the infinite product
looks like a thermal distribution function obeying Bose
statistics. Going back to the absorption probability
given by the second line of Eq.~(\ref{Aprob}), the factor $\sinh\pi k$ 
there is canceled by the above $1/\sinh\pi k$ factor, leaving
only a finite number of zeros at $ik=1,2,\cdots,m$.
We speculate that the original factor $\sinh\pi k$ in the second
line of Eq.~(\ref{Aprob}) is something to do with the nature of the wall,
while the $1/\sinh\pi k$ factor in Eq.~(\ref{IP:even})
indeed represents the thermal spectrum due to the acceleration of 
the boundary wall: Since the boundary wall is a dS spacetime with dS
temperature $T_{\rm dS}=(2\pi a_w )^{-1}$, which can be also regarded
as the Unruh temperature due to its acceleration relative to the bulk,
observers on the boundary wall see a thermal spectrum of gravitons.
This thermal spectrum is known to be in the form of Bose statistics
in even spacetime dimensions, i.e., for odd $d$,
while in the form of Fermi statistics in odd spacetime dimensions, i.e.,
for even $d$~\cite{Ooguri:1985nv}.
The above argument is in fact supported if we consider
the case of odd spacetime dimensions, which we now turn to.

In the case of odd spacetime dimensions (or even $d$) equal to
or more than five, $d=2m+4$ or $\nu=m+3/2$ ($m=0,1,2,\cdots$),
Eq.~(\ref{Aprob}) gives
\begin{eqnarray}
{\cal A} (k) &=& \frac{\pi k\tanh \pi k }{2^{4m+3}}
\left(\frac{\ell_{\rm AdS}}{a_w}\right)^{2m+3}
\left[ \frac{(2m+1)!!}{(m+1)!}\right]^2 
\prod\limits^{m}_{n=0} 
\left( 1+\frac{4k^2}{\left(2n+1\right)^2 }\right) 
\bigg|F_{m+1/2,k}(x_0')\bigg|^{-2} \,,
\label{AB:d=2m+4}
\end{eqnarray}
where we have used
\begin{eqnarray}
\left[ \prod\limits^{\infty}_{n=1}
\left(1+\frac{4k^2}{\left(2m+1+2n\right)^2}
\right) \right]^{-1}
=\prod\limits^{m}_{n=0} 
\left(1+\frac{4k^2}{(2n+1)^2}
\right)
\frac{1}{\cosh\pi k}  \,.
\label{IP:odd}
\end{eqnarray}
We find that zeros due to the factor $\sinh\pi k$ in Eq.~(\ref{Aprob}) 
remain and consequently poles in the form 
$\tanh\pi k$ due to the factor $1/\cosh\pi k$ in the 
infinite product in Eq.~(\ref{IP:odd}) appear.
Thus, an infinite number of poles which are not
related to graviton bound states show up 
in odd spacetime dimensions. The factor $1/\cosh\pi k$
represents a thermal spectrum of Fermi statistics.
This difference from even spacetime dimensions
can be attributed to the absence of Huygens principle
in odd spacetime dimensions. As mentioned in the case of
even spacetime dimensions, the above analysis 
is consistent with the fact that an accelerated observer 
see a thermal spectrum at Unruh temperature, with the 
spectrum obeying Bose (Fermi) statistics in even (odd)
spacetime dimensions.

Now, we try to express the absorption probability in an analogous form
to Eq.~(\ref{A:4DCFT}).

In the case of even spacetime dimensions (or odd $d$) equal to
or more than four,  $d=2m+3$ ($m=0,1,2,\cdots$), Eq.~(\ref{AB:d=2m+3}) 
may be expressed in the form,
\begin{eqnarray}
{\cal A} (k) =  c_{2m+4} \ {\cal G}_{2m+4} \ k_w^2~ \prod\limits^{m}_{n=1} 
\left( n^2\lambda_{\rm IR}^2 + k_w^2\right) \, .
\label{A:d=2m+3}
\end{eqnarray}
Here, we have defined the 
$(2m+4)$-dimensional degree of freedom $c_{2m+4}$ of the CFT, 
and the $(2m+4)$-dimensional effective gravitational coupling constant, 
${\cal G}_{2m+4}$, which are expressed as
\begin{eqnarray}
&&c_{2m+4} = \left[ \frac{2}{(2m+1)!!}\right]^2
\frac{\ell_{\rm AdS}^{2m+2}}{G_{2m+4} }
\ ,
\nonumber\\
&&{\cal G}_{2m+4} 
= G_{2m+4}\,\bigg| F_{m,k}(x_0')\bigg|^{-2}\,.
\label{CG:d=2m+3}
\end{eqnarray} 
We note that $c_{2m+4}$ stems from the AdS/CFT correspondence and
${\cal G}_{2m+4}$ includes the information of the dS bulk geometry, 
which is different from the gravitational coupling constant 
in the dS bulk $G_{2m+4}$. 
As we will see later in subsection~\ref{subsec:genericcases},
the difference between 
${\cal G}_{2m+4}$ and $G_{2m+4}$ may be attributed 
to the effect of localization of gravitons in the dS bulk.

In the case of odd spacetime dimensions (or even $d$) equal to
or more than five, $d=2m+4$ ($m=0,1,2,\cdots$), Eq.~(\ref{AB:d=2m+4}) gives
\begin{eqnarray}
{\cal A} (k) =  c_{2m+5}~ {\cal G}_{2m+5}~ 
k_w\tanh \pi k~
\prod\limits^{m}_{n=0} 
\left( \frac{\left(2n+1\right)^2}{4}\lambda_{\rm IR}^2 + k_w^2\right) 
 \,,
\label{A:d=2m+4}
\end{eqnarray}
where we have defined 
\begin{eqnarray}
&&c_{2m+5} = \frac{\pi}{2^{2m+1}} \left[ \frac{1}{(m+1)!}\right]^2
\frac{\ell_{\rm AdS}^{2m+3}}{G_{2m+5}}
\,,
\nonumber\\
&&{\cal G}_{2m+5} = G_{2m+5}~ 
\bigg|F_{m+1/2,k}(x_0')\bigg|^{-2} \, .
\label{CG:d=2m+4}
\end{eqnarray}

As we noted earlier below Eq.~(\ref{M1}), 
the factorized dependence on the dS bulk ($x_0'$ or 
$\ell_{\rm dS}$) and the AdS spacetime ($x_0$ or $\ell_{\rm AdS}$)
is encoded in ${\cal G}_{d+1}$ and $c_{d+1}$ in the CFT language, 
respectively.
In fact, if we consider the AdS bubble in the Minkowski spacetime,
we find there is no correction to the effective gravitational
constant, ${\cal G}_{d+1}=G_{d+1}$.\footnote{As we will see below
in Section~\ref{subsec:Minbulk}, however, if we first consider
a dS bulk and then take the Minkowski bulk limit,
this statement does not hold in the case of odd spacetime dimensions
(even $d$). Regrettably we do not have a convincing explanation
for this discrepancy.} This is because we took the decoupling limit.

We now look at the absorption probability in three special
cases in which the above analytic formula can be simplified:
\begin{itemize}
\item Minkowski disc in AdS spacetime:
\\\quad
$\ell_{\rm dS}/a_w=\cosh z_0' \to\infty$, 
or $x_0'=\tanh z_0'=1$.
\item AdS bubble in the Minkowski spacetime:
\\\quad
$\ell_{\rm dS}/a_w=\cosh z_0' \to\infty$, or $x_0'=\tanh z_0'=-1$.
\item The largest possible bubbles in the dS bulk:
\\\quad
$\ell_{\rm dS}=a_w\cosh z_0' =a_w$, or $x_0'=0$.
\end{itemize}
More generic cases can be investigated perturbatively and numerically.
We discuss the above three cases separately.

\subsection{Minkowski disc in AdS spacetime}
\label{subsec:Mindisc}

We consider the case of a Minkowski disc surrounded by the domain wall 
with radius $a_w$.
That is, $\ell_{\rm dS}/a_w \to\infty$ and $z_0'\to+\infty$. 
From Eqs.~(\ref{continuity1}) and (\ref{x0x0pdef}), this corresponds to
\begin{eqnarray}
x_0^\prime=\tanh z_0'=1\,.
\label{lowenergy1}
\end{eqnarray} 
In this case, 
the effective gravitational coupling ${\cal G}_{d+1}$ becomes $G_{d+1}$
because of $F(a,b,c;0)=1$, where $a,b$ and $c$ are arbitrary real or complex 
parameters. 

In the case of even spacetime dimensions, $d=2m+3$, 
we find 
\begin{eqnarray}
{\cal A} (k) = c_{2m+4}~G_{2m+4}~k_w^2   
~\prod\limits^{m}_{n=1} 
\left( n^2\lambda_{\rm IR}^2 + k_w^2\right) \ .
\label{evenMdisc1}
\end{eqnarray}
We can easily check that Eq.~(\ref{A:4DCFT}) can be reproduced 
from this result in the case of four dimensions $(d=3,~\mbox{or}~m=0)$.
We find corrections caused by the curvature radius of the domain wall
$\lambda_{\rm IR}^2= 1/a_w^2$ come into the result. 
However, taking the flat wall limit:
$a_w \cdot  k_w \to\infty$ with $k_w$ fixed,
we obtain the standard result in the Minkowski spacetime. 
In the case of six dimensions $(d=5,~\mbox{or}~m=1)$, Eq.~(\ref{evenMdisc}) 
becomes
\begin{eqnarray}
{\cal A}(k)&=&c_{6} \  G_6 \ k_w^2 \left(\lambda_{\rm IR}^2 + k_w^2
\right)
\nonumber\\
&&\hspace{-5mm}
\xrightarrow{\rm flat\, wall}~
c_6 \ G_6 \left(\Delta E\right)^4\,.
\label{trans:six}
\end{eqnarray}

In the case of odd spacetime dimensions, $d=2m+4$, 
Eq.~(\ref{A:d=2m+4}) gives
\begin{eqnarray}
{\cal A}(k)= c_{2m+5}~ G_{2m+5}~ k_w \tanh \pi k  
~\prod\limits^{m}_{n=0} 
\left( \frac{\left(2n+1\right)^2}{4}\lambda_{\rm IR}^2 + k_w^2\right)  \ .
\label{oddMdisc1}
\end{eqnarray}
We see that there are corrections due to the curvature radius of
the domain wall $\lambda_{\rm IR}^2$ again and also the thermal nature
 of the wall $\tanh \pi k$ appears in the result. 
However, if we take the flat wall limit, then $\tanh \pi k\to 1$.
So all the corrections disappear in this limit.

In particular in the case of five dimensions $(d=4,~\mbox{or}~m=0)$, we have
\begin{eqnarray}
{\cal A}(k)&=&c_5~ G_5~ k_w \tanh \pi k~
\left(\frac{1}{4}\lambda_{\rm IR}^2 + k_w^2 \right)
\nonumber\\
&&\hspace{-5mm}
\xrightarrow{\rm flat\, wall}~
c_5 \ G_5 \left(\Delta E\right)^3  \ .
\label{4dimtrans}
\end{eqnarray}

\subsection{AdS bubble in Minkowski spacetime}
\label{subsec:Minbulk}

We now consider an AdS bubble in Minkowski spacetime.
That is, $\ell_{\rm dS}/a_w \to\infty$ and $z_0'\to-\infty$. 
From Eqs.~(\ref{continuity1}) and (\ref{x0x0pdef}), this corresponds to
\begin{eqnarray}
x_0^\prime=\tanh z_0'=-1\,.
\end{eqnarray} 

In fact, in the case of even spacetime dimensions, $d=2m+3$, 
we find the hypergeometric function in Eq.~(\ref{CG:d=2m+3}) 
is still unity,
$
|F_{m,k}(-1)|=|F(-m,m+1,1-ik,1-0)|=1\,.
$
Hence we obtain exactly the same result as in the case of
a Minkowski disc in AdS spacetime,
\begin{eqnarray}
{\cal A} (k) = c_{2m+4}~ G_{2m+4}~ k_w^2   
~\prod\limits^{m}_{n=1} 
\left( n^2\lambda_{\rm IR}^2 + k_w^2\right) \ .
\label{evenMdisc2}
\end{eqnarray}
In the case of four dimension $(d=3,~\mbox{or}~m=0)$, we get the same 
result as Eq.~(\ref{A:4DCFT}). Also, we have 
Eq.~(\ref{trans:six}) in six dimension $(d=5,~\mbox{or}~m=1)$. 
Thus the standard result in the Minkowski spacetime is 
also applied to this case in
the flat wall limit.

However, in the case of odd spacetime dimensions, $d=2m+4$,
the situation is a bit more complicated.
The value of the hypergeometric function in Eq.~(\ref{A:d=2m+4})
in the limit $x_0'\to-1$ takes the form,
\begin{eqnarray}
F_{m+1/2,k}(x_0')\mathop{\longrightarrow}\limits_{x_0'\to-1}
F\left(-m-\frac{1}{2},~m+\frac{3}{2},~1-ik;~1-0\right)
=\frac{e^{i\theta(k)}}{\sinh \pi k}
\left(\cosh \pi k +e^{2ikz_0'}e^{i\varphi(k)}\right)\,,
\end{eqnarray}
where the phases $\theta(k)$ and $\varphi(k)$ depend
on $k$ in a complicated way. But the important point is
the phase factor $e^{ikz_0'}$. It oscillates indefinitely
in the limit $z_0'\to-\infty$. 
This seems to imply the loss of rigorous information. 
Namely, one may not be able to recover the whole information
of the bulk in this case. In any case, the result is
\begin{eqnarray}
\bigg|F\left(-m-\frac{1}{2},~m+\frac{3}{2},~1-ik;~1-0\right)\bigg|^{-2}
=\frac{\sinh^2\pi k}{\cosh^2\pi k+1}\,.
\end{eqnarray}
Thus, the effective gravitational coupling in Eq.~(\ref{CG:d=2m+4}) 
includes the effect of the dS bulk such as
\begin{eqnarray}
{\cal G}_{2m+5} = G_{2m+5}~ \frac{\sinh^2\pi k}{\cosh^2\pi k+1} \, .
\label{effGMin_odd}
\end{eqnarray}
It turns out that the ${\cal G}_{2m+5}$ is smaller than $G_{2m+5}$.

The absorption probability is given by
\begin{eqnarray}
{\cal A}(k) = c_{2m+5}~ {\cal G}_{2m+5}~k_w \tanh \pi k
~\prod\limits^{m}_{n=0} 
\left( \frac{\left(2n+1\right)^2}{4}\lambda_{\rm IR}^2 + k_w^2\right)\,.
\label{oddMdisc2}
\end{eqnarray}
Compared to the case of the Minkowski disc, Eq.~(\ref{oddMdisc1}),
we see that the absorption probability is smaller in the present case
due to the suppressed effective gravitational constant~(\ref{effGMin_odd}).
We found corrections coming from the curvature radius of the domain wall
$\lambda_{\rm IR}^2$ and also the thermal nature of the dS bulk
$\tanh \pi k$ in the result. 
However, taking the flat wall limit:
$a_w \cdot k_w \to\infty$ with $k_w$ fixed,
then $\tanh \pi k\to 1,\,\sinh^2 \pi k/(\cosh^2 \pi k +1)\to 1$. 
Again, all the corrections disappear in this limit.

In particular in the case of five dimensions $(d=4,~\mbox{or}~m=0)$,
we obtain
\begin{eqnarray}
{\cal A}(k)&=& c_5~{\cal G}_{5}~k_w \tanh \pi k~ 
\left(\frac{1}{4}\lambda_{\rm IR}^2+k_w^2\right) 
\nonumber\\
&&\hspace{-5mm}
\xrightarrow{\rm flat\, wall}~
c_5 ~G_5 \left(\Delta E\right)^3  \, .
\label{4.29}
\end{eqnarray}

\subsection{Maximal radius bubbles in dS universe}
\label{subsec:maxbubble}

Now, we consider the largest possible bubbles in the dS bulk.
That is, $\ell_{\rm dS}=a_w$ or $z_0'=0$. 
From Eq.~(\ref{x0x0pdef}), this corresponds to
\begin{eqnarray}
x_0^\prime=\tanh z_0'=0\,.
\end{eqnarray} 
In this limit, the hypergeometric
function $F_{\nu-1,k}(x_0')$ in the 
formulas~(\ref{CG:d=2m+3}) and (\ref{CG:d=2m+4})
can be evaluated by using the formula,
\begin{eqnarray}
F\left(-a,~1+a;~b;~\frac{1}{2}\right)
=\frac{2^{1-b}\sqrt{\pi}\,\Gamma(b)}
{\Gamma\left(\dfrac{b-a}{2}\right)\Gamma\left(\dfrac{b+a+1}{2}\right)}\,.
\label{hgformula}
\end{eqnarray}

For even spacetime dimensions, $d=2m+3$, for which $a=m$, 
the above can be explicitly evaluated in terms of elementary functions.
The result is
\begin{eqnarray}
\left|F_{m,k}(0)\right|^{-2}
=\left|F\left(-m,~1+m;~1-ik;~\frac{1}{2}\right)\right|^{-2}
=\left\{
\begin{array}{ll}
\prod\limits_{\ell=1}^{p}
\dfrac{4\ell^2+k^2}{(2\ell-1)^2+k^2}\,;
&~m=2p,\hspace{7mm}p=0,1,2\cdots
\\
~\\
\prod\limits_{\ell=0}^{p}
\dfrac{(2\ell+1)^2+k^2}{4\ell^2+k^2}\,;
&~m=2p+1
\end{array}
\right.
\label{evenfactor}
\end{eqnarray}
Thus, the effective gravitational coupling in even spacetime dimensions 
in Eq.~(\ref{CG:d=2m+3}) is enhanced when $m=2p$ as
\begin{eqnarray}
{\cal G}_{2m+4} =  G_{2m+4}~  \prod\limits_{\ell=1}^{p}
\frac{4\ell^2+k^2}{(2\ell-1)^2+k^2} \,,
\label{max:even}
\end{eqnarray}
and when $m=2p+1$ as
\begin{eqnarray}
{\cal G}_{2m+4} = G_{2m+4}~
\prod\limits_{\ell=0}^{p}
\dfrac{(2\ell+1)^2+k^2}{4\ell^2+k^2} \,.
\label{max:odd}
\end{eqnarray}
We find that ${\cal G}_{2m+4}$ is always greater than $G_{2m+4}$ for
both cases. Hence, the form of the absorption
probability is the same as in the previous two Minkowski cases,
but it is enhanced compared to them due to the enhancement in
the effective gravitational constant.
We again find corrections caused by the curvature radius of the domain 
wall, which vanish in the flat wall limit $k=a_w\,k_w\to\infty$ with
$k_w$ fixed. 
Note that we have poles in Eqs.~(\ref{max:even}) and (\ref{max:odd}), 
which are related to the bound states discussed in Appendix~\ref{app:BS},
 Eq.~(\ref{boundstate}).

For four dimensions ($d=3$ or $m=0$), Eq.~(\ref{A:4DCFT}) 
is recovered from the above result because $F_{0,k}(x_0')=1$ 
independent of $x_0'$ as we mentioned earlier.

In the case of six dimensions ($d=5$ or $m=1$), the effect of
the dS bulk is encoded in ${\cal G}_{6}$ in Eq.~(\ref{max:odd}).
So from Eq.~(\ref{A:d=2m+3}) we obtain a slightly enhanced absorption
probability compared to the previous two Minkowski cases,
\begin{eqnarray}
{\cal A}(k) &=& c_6~{\cal G}_{6}~k_w^2~\left(\lambda_{\rm IR}^2 + k_w^2
\right)
=c_6~G_6~\left(\lambda_{\rm IR}^2 + k_w^2\right)^2
\nonumber\\
&&\hspace{-5mm}
\xrightarrow{\rm flat\, wall}~
c_6 ~G_6 (\Delta E)^4  \,.
\label{transmission4}
\end{eqnarray}

In the case of odd dimensions, $d=2m+4$, we put $a=m+1/2$
in the formula (\ref{hgformula}). In this case there does not
seem to exist a way to express the hypergeometric factor in
terms of elementary functions. So we are unable to express
the effective gravitational coupling ${\cal G}_{2m+5}$ in a simple form.
Nevertheless, we can estimate whether it is possible
to describe the result in terms of the CFT language or not by 
considering the product of
the absorption probability of two adjacent odd spacetime dimensions.

For definiteness, let us consider
five dimensions $(d=4)$ and seven dimensions $(d=6)$. 
Here, as there are no gravitational waves
in three dimensions $(d=2)$, we have skipped this case.
If the result could be described by the CFT language, we expect the
product of them should also be described by it.
 The result we obtain from Eq.~(\ref{A:d=2m+4}) is
\begin{eqnarray}
{\cal A}_{d=4}\times{\cal A}_{d=6}
&=& c_5~{\cal G}_5~ c_7~{\cal G}_7~
k_w^2\tanh^2 \pi k~\left(\frac{1}{4}\lambda_{\rm IR}^2+k_w^2\right)^2
\left(\frac{9}{4}\lambda_{\rm IR}^2+k_w^2\right)
\nonumber\\
&=& c_5~G_5~ c_7~G_7~\tanh^4 \pi k
\left(\frac{1}{4}\lambda_{\rm IR}^2 + k_w^2\right)^2
\left(\frac{9}{4}\lambda_{\rm IR}^2 + k_w^2\right)^2
\nonumber\\
&&\hspace{-5mm}
\xrightarrow{\rm flat\, wall}~
c_5~G_5(\Delta E)^3\cdot
c_7~G_7(\Delta E)^5 
\,.\label{transmission5}
\end{eqnarray}
Here again, we have corrections due to the curvature radius of 
the domain wall $\lambda_{\rm IR}^2$ and also to
the thermal nature of the domain wall. 
Note, however, there is an additional factor $\tanh^2\pi k$ in the result.
This may be due to the thermal nature of the dS bulk that was absent
in the Minkowski bulk limit.

If we take the flat wall limit: $k=a_w \, k_w  \to\infty$ with 
$k_w$ fixed, all the corrections again disappear
and the result is written by the product form 
of each absorption probability in five and seven dimensions.
Thus we expect that 
the standard result in the Minkowski spacetime is recovered 
in the flat wall limit.

\subsection{Generic cases}
\label{subsec:genericcases}

In four dimensions, we did not see any effect of corrections to 
the decay rate. This is probably related to the fact that 
three-dimensional gravity on the domain wall is trivial.
In higher dimensions, we found that the dependence on $a_w$ appears in 
the absorption probability explicitly, which vanishes in the flat 
wall limit: $k=a_w \, k_w \to\infty$ with $k_w$ fixed,
in all the cases studied in the previous subsections. 

In the present model, we took the decoupling limit in the AdS side.
Hence, the CFT would not couple to gravity on the domain wall 
if there were no dS bulk beyond the wall. 
Here, however, there is a dS bulk outside the AdS bubble and
there are localized gravitons in the dS bulk (see Appendix~\ref{app:BS}). 
Thus, we intuitively expect that not only corrections due to
the curvature of the domain wall $a_w$ but also
effects of localized gravitons in the dS bulk with $\ell_{\rm dS}$ 
would come in to the absorption probability. 
We studied analytically the cases of the Minkowski disc,
the Minkowski bulk and the maximal radius bubble.
In the former two cases there remains
no effect of $\ell_{\rm dS}$ because the limit
$\ell_{\rm dS}\to\infty$ is taken, and in the
last case of the maximal radius bubble in the dS bulk,
we cannot clearly see effects of $\ell_{\rm dS}$ either
because the bubble radius and the dS bulk curvature radius is
degenerate. This is the reason why we could see only the 
corrections due to $a_w$ in the previous subsections.

In order to reveal the effect of the dS bulk on the absorption 
probability we consider more generic cases in this subsection.

\begin{figure}[ht]
\includegraphics[width=9cm]{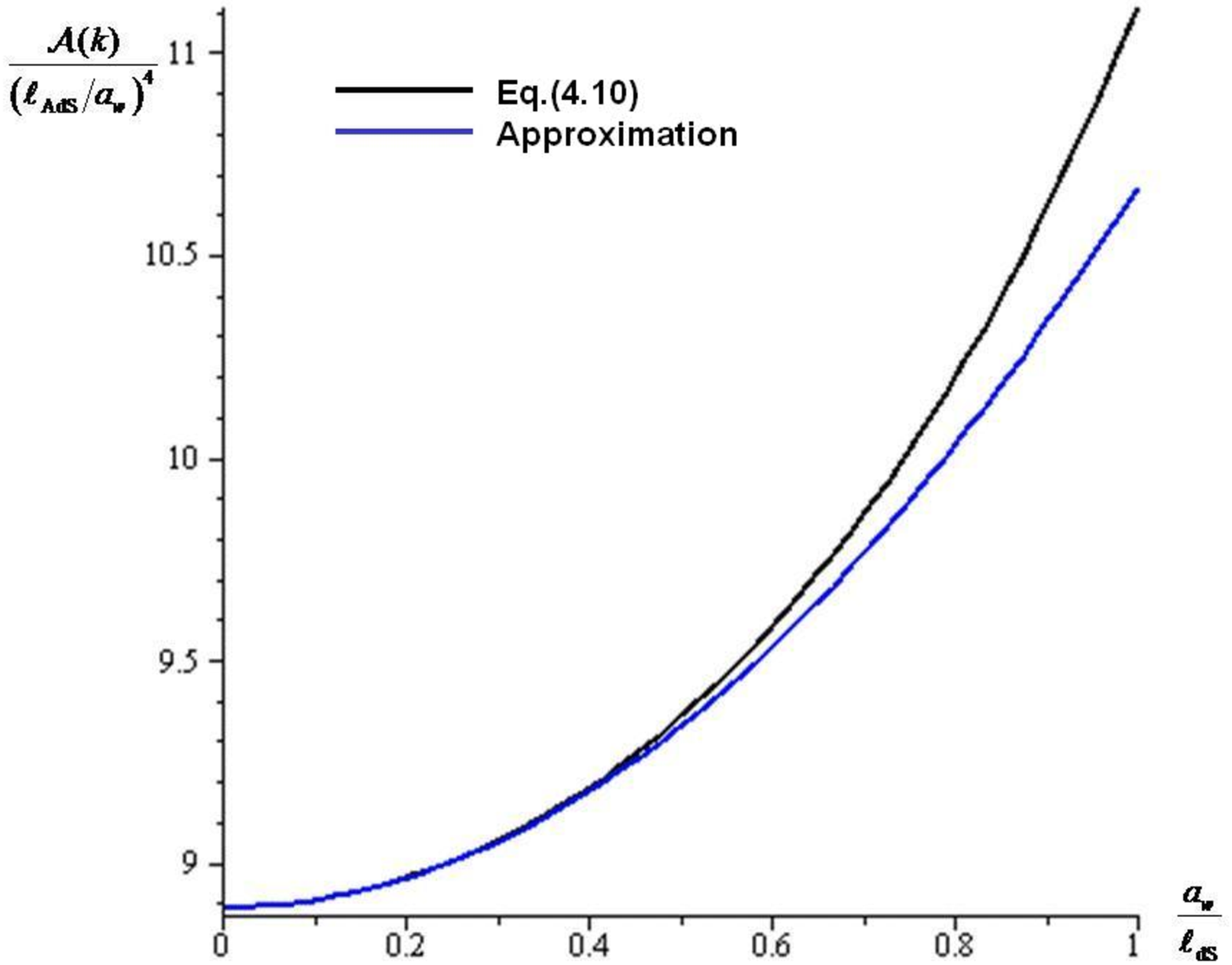}
\caption{The absorption probability is plotted as a function of 
$a_w/\ell_{\rm dS}$.}
\label{fig:4}
\end{figure}

We first study the effect of dS bulk in the vicinity of
the Minkowski disc limit.
By expanding the hypergeometric function around $x_0^\prime=1$,
we see leading corrections due to finite curvature radius of
 the dS bulk to the effective gravitational coupling constant.
In the case of even spacetime dimensions more than three, $d=2m+3$,
we have
\begin{eqnarray}
{\cal G}_{2m+4} = G_{2m+4}\left[ 1 + \frac{m(m+1)}{2(1+k^2)} 
\left(\frac{a_w}{\ell_{\rm dS}}\right)^2 \right] 
\,.
\label{d=2m+3}
\end{eqnarray}
We see that the effect is to enhance the gravitational constant
in general, except for four dimensions.
We find there is no correction in four dimensions ($m=0$).
In the case of odd spacetime dimensions more than four, 
$d=2m+4$, we have
\begin{eqnarray}
{\cal G}_{2m+5} = G_{2m+5}
\left[ 1 + \frac{(2m+1)(2m+3)}{4(1+k^2)} 
\left(\frac{a_w}{\ell_{\rm dS}}\right)^2 \right] \,.
\label{d=2m+4}
\end{eqnarray}
Also in this case 
we find the effect of finite $a_w/\ell_{\rm dS}$ is to
increase the strength of the effective gravitational coupling.
In Fig.~\ref{fig:4}, 
we depict ${\cal A}(k)/(\ell_{\rm AdS}/a_w)^4$ in Eq.~(\ref{AB:d=2m+3})
and approximated one in Eq.~(\ref{d=2m+3}) 
with fixed $k=2$ in six dimensions $(d=5,~\mbox{or}~m=1)$
as a function of $a_w/\ell_{\rm dS}$. We find the approximation is good
enough.

We see that the absorption probability approaches Eq.~(\ref{trans:six}) as
one approaches the Minkowski disk (to the left) and 
also approaches Eq.~(\ref{transmission4}) 
as one moves toward the Maximal radius bubble (to the right). 
Thus the effect of the non-vanishing dS curvature increases 
the absorption probability. It is also easy to perform
the perturbative analysis around the limit of AdS bubble in
Minkowski spacetime $x_0' =-1$.
A similar calculation leads to the same result as
given by Eqs.~(\ref{d=2m+3}) and (\ref{d=2m+4}).
Thus the behavior of the absorption probability near the
Minkowski limit or for fairly large $k$
can be understood from the perturbative result
for the gravitational coupling constant.  

\begin{figure}[htbp]
\includegraphics[width=10cm]{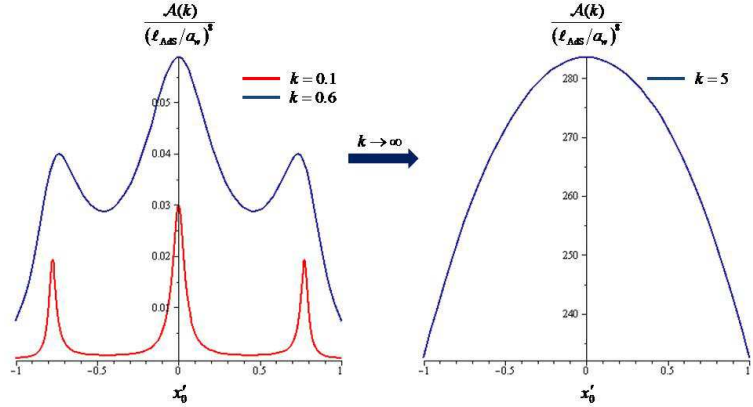}
\vspace{-1cm}
\caption{The absorption probability for ten dimensions
 ($d=9$ or $m=3$) is plotted.}
\label{fig:5}
\end{figure}
\begin{figure}
\includegraphics[width=10cm]{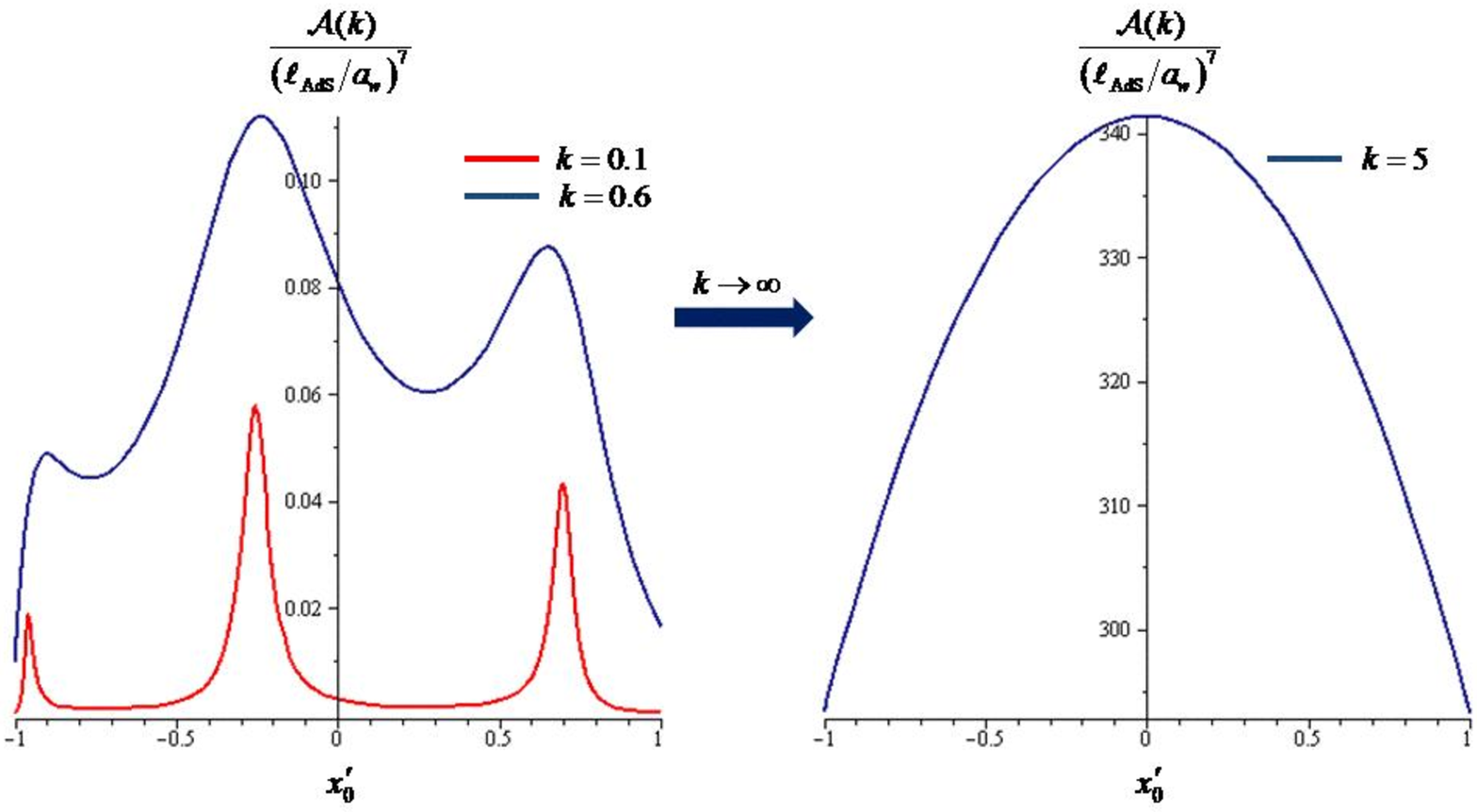}
\vspace{-1cm}
\caption{The absorption probability for nine dimensions $(d=8,~\mbox{or}~m=2)$
is plotted.}
\label{fig:6}
\end{figure}

Although the approximation $F(a,~b,~1-ik;~z)\approx 1+\frac{ab}{1-ik}z$
is valid for large $k$, it is not going to be valid for small $k$.
In fact, for $k\leq 1$, we find qualitatively different behavior.
In Figs.~\ref{fig:5} and \ref{fig:6}, we plot the absorption probability
for ten dimensions ($d=9$ or $m=3$) and nine dimensions 
($d=8$ or $m=2$) at $k=0.1$, $0.6$ and $k=5$. 
We clearly see resonant peaks in each figure when $k$ is small enough. 
The number of the peaks is found to coincide with 
the number of massive graviton bound states in the limit of
Minkowski bulk as discussed in Appendix~\ref{BSMinbulk}.

In the case of even dimensions, Fig.~\ref{fig:5}, 
the resonance is symmetrical with respect to $x_0'=0$ and 
it is maximum at $x_0'=0$ irrespective of $k$. 
On the other hand for odd dimensions, Fig.~\ref{fig:6}, 
the symmetry disappears and the maximum peak shifts toward 
a negative $x_0'$. In both even and odd dimensions,
the resonance gradually becomes less prominent as $k$ increases,
and merges into a single peak around $x_0'=0$ at $k\gg1$.

This resonance seems to occur due to massive localized gravitons 
in the dS bulk. We find that the number of massive bound states of 
gravitons increases as the position of the domain wall moves 
from $x'_0=1$ to $-1$. We confirmed this behavior
by numerical calculations. More precisely, 
the position of each peak in Fig.~\ref{fig:5} or Fig.~\ref{fig:6} 
corresponds to the appearance of a new massive bound state 
as we move the domain wall from $x'_0=1$ to $-1$. 

Let us try to explain this behavior by using Fig.~\ref{fig:5} 
with the help of Figs.~\ref{fig:2} and \ref{fig:3}. 
In the Minkowski disc limit $x'_0=1$, there is no massive
bound state as discussed in Appendix~\ref{BSMindisc}. 
The potential is flat in this limit, so the resonance does not occur.
As the domain wall moves to the left from the limit, 
the potential well starts to appear and a bound state appear at
some point. The energy of the bound state when it appears
is $k^2=0$ and gradually it becomes $k^2<0$. Hence  
for incoming gravitons $k$ whose energy is sufficiently
small, $k\ll1$, the resonance happens.

As the domain wall moves further towards $x'_0=0$, the potential
well gets deep and the energy of the massive bound state decreases.
Then the resonance no longer happens. However, as the potential becomes 
deeper a new massive bound state may appear and the resonance
happens again. This makes the second peak in Fig.~\ref{fig:5}.
In this case of ten dimensions, the second peak happens to
be at the symmetric point $x_0'=0$. 

Next, as the domain wall goes away further from $x'_0=0$ to negative
$x_0'$, the area of the potential well increases and the energy of 
the massive bound stated of gravitons decreases.
 Then the resonance ceases to happen.
However, as the area of the potential well increases
as the domain wall moves toward to the Minkowski limit $x'_0=-1$, 
there may appear another bound state and the resonance may
occur again. This is in fact the case in ten dimensions.
As the area of the potential further increases, this resonance
ceases to happen again, and after the area of
the potential approaches exponentially to the asymptotic value,
the number of bound states no more changes.
The absorption probability decreases and approaches
a finite value asymptotically in the limit $x_0'\to-1$.
These features are encoded in the effective gravitational
coupling ${\cal G}_{d+1}$.

\subsection{dS/dS and dS/CFT correspondence}
\label{subsec:dsds}

\begin{figure}[ht]
\begin{center}
\begin{minipage}{8cm}
\begin{center}
\hspace{-1.5cm}
\includegraphics[width=65mm]{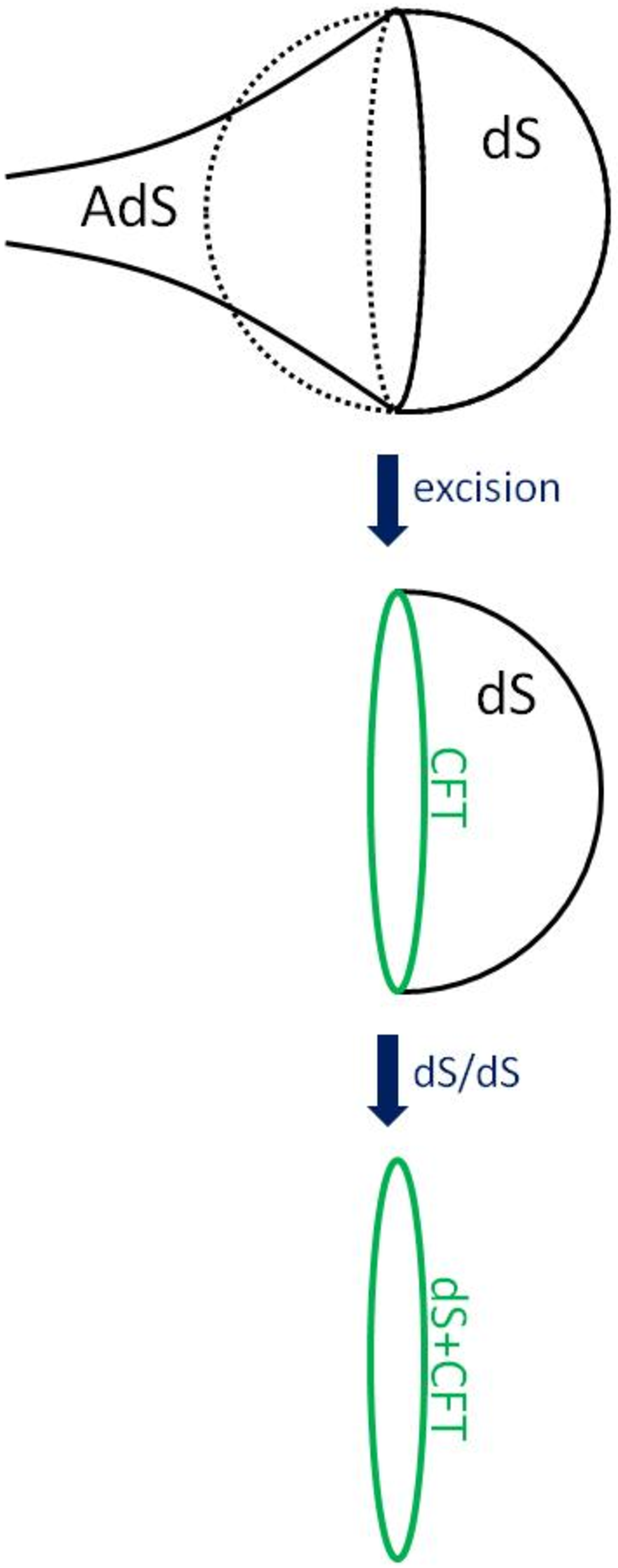}\vspace{0cm}
\caption{The dS spacetime 
with the boundary CFT after excision of the AdS spacetime
is also mapped onto the domain wall by the dS/dS correspondence.}
\label{fig:7}
\end{center}
\end{minipage}
\hspace{1mm}
\begin{minipage}{8cm}
\begin{center}
\hspace{-1.5cm}
\includegraphics[width=65mm]{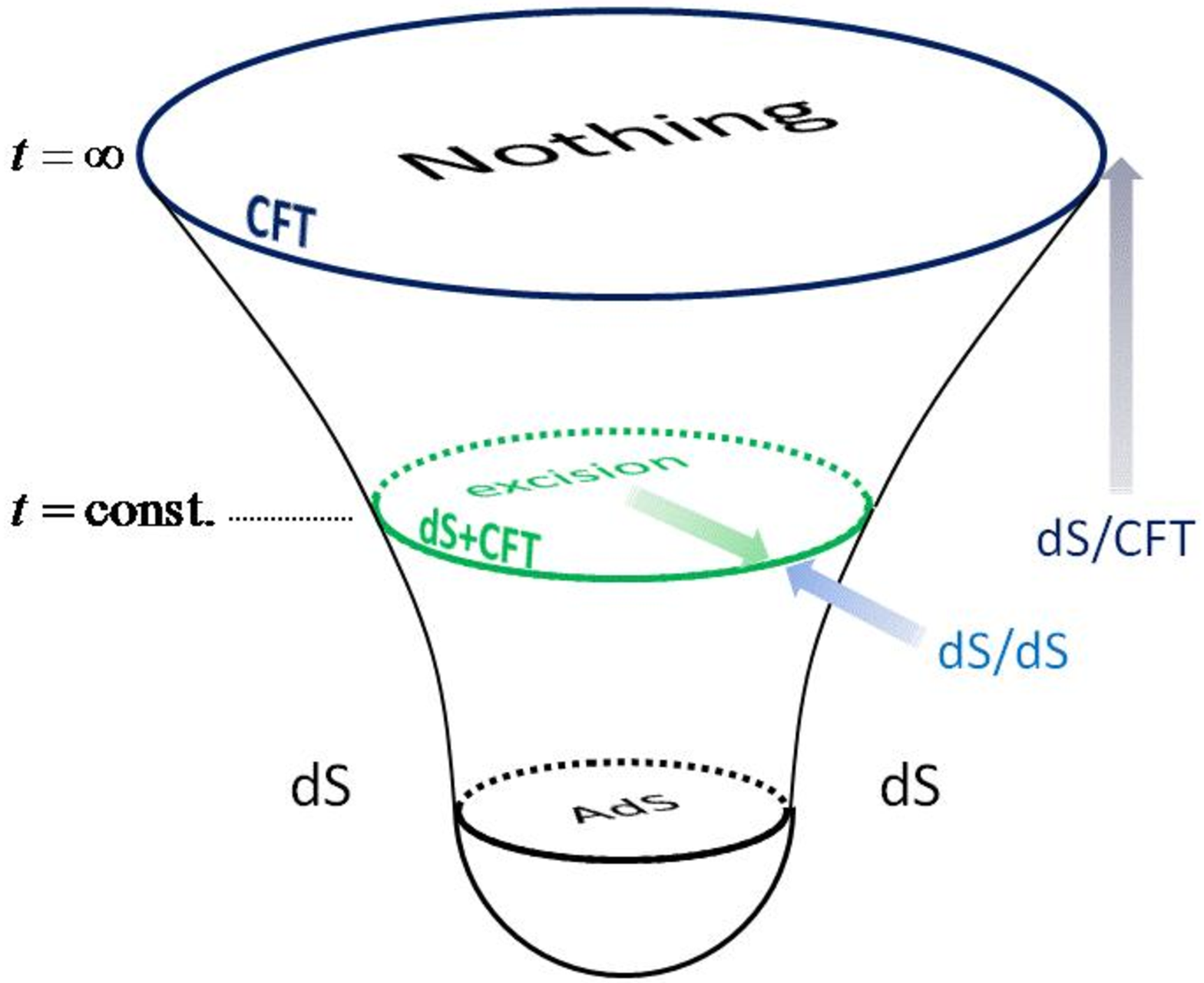}\vspace{10mm}
\caption{A bubble evolution is depicted. After the mapping by
 the dS/dS correspondence,
neither inside nor outside of the bubble wall exists. 
There exists a CFT coupled to gravity. Furthermore, we could map this
onto a $(d-1)$-dimensional CFT in the future boundary of
$d$-dimensional dS according to the dS/CFT correspondence.}
\label{fig:8}
\end{center}
\end{minipage}
\end{center}
\end{figure}

We have considered the dual of the dS universe with an AdS bubble,
that is, the dS bulk with the boundary on which CFT resides.
In this dual picture, as the AdS spacetime disappears after 
excision, we expect that the energy of gravitons coming from the 
dS bulk should be absorbed by CFT matter.
We confirmed that the result for all the cases 
studied in the previous subsections was in agreement
with the standard form of decay rate of gravitons to CFT
particles in the flat wall limit. 
Through the calculation of absorption probability, 
we saw that the effect of AdS spacetime comes in only through 
effective degrees of freedom $c_{d+1}$ of CFT matter.
In this absorption process, we found that the effective gravitational 
coupling  ${\cal G}_{d+1}$ between gravitons and CFT matter
is controlled by localized gravitons.
As shown in Appendix~\ref{app:BS}, the number of localized 
gravitons depends on the dimensions and the position of the domain wall. 
We found that for $k\gtrsim1$
the absorption probability becomes maximal at the maximal 
radius bubble because the effective gravitational coupling becomes 
maximum there. Moreover, since the massive bound states exist, 
the absorption probability shows resonant peaks at low energy, $k\ll1$. 
The number of resonant peaks is found to be 
the same as the number of the bound states in the Minkowski bulk limit.
For incident gravitons with sufficiently small energy,
the resonance with a bound state occurs when it appears at $k^2=0$.
This leads to the sharp resonant decay of gravitons
into CFT matter. 
This happens at the position where a new bound state appears.

We can further think of dual of the dS bulk with
the boundary CFT, a la dS/dS 
correspondence~\cite{Karch:2003em, Alishahiha:2004md, Alishahiha:2005dj},
although our model is slightly different from the exact 
dS/dS correspondence in that there is an AdS bubble
connected to the dS bulk through a domain wall.
In fact, the existence of localized (massless as well as massive)
gravitons in the $d+1$ dS bulk guarantees the existence of 
the $d$-dimensional Newton constant. Thus, we can map 
$(d+1)$-dimensional gravity onto $d$-dimensional gravity by
integrating out an extra dimension where the localized gravitons exist.
Then propagating gravitons coming from the $(d+1)$-dimensional dS bulk
 turn out to be KK gravitons on the one lower dimensional domain wall. 
In this dS/dS correspondence picture, 
KK gravitons could decay into CFT matter in two ways:
One is the direct decay of KK gravitons into CFT matter.
The other is the indirect decay through massive gravitons,
namely, KK gravitons decay first to massive gravitons, 
and they eventually decay into CFT matter.
Therefore the effective gravitational coupling constant between 
KK gravitons and CFT matter is enhanced by the second 
process.

Also, as CFT matter lives on the domain wall, the decay probability
 depends on the dS radius on the wall $a_w$. 
The main difference from the usual dS/dS correspondence
 is that the position of the domain wall $x_0'$ 
changes the shape of potential as we see in Fig.~\ref{fig:2} and 
Fig.~\ref{fig:3}. 
Then the mass spectrum of the localized gravitons 
depends on the position of the domain wall. As a consequence, 
the decay probability also depends on $x_0'$ or equivalently on 
$\ell_{\rm dS}$. This is also expressed in terms of
the effective gravitational coupling constant
 where the poles in the complex $k$-plane corresponding to the 
graviton bound states show up.
In other words, (dS with boundary)/dS correspondence can be parametrized
by the position of the domain wall $x_0'$. The dependence of the decay
rate on $x_0'$ is encoded in the effective gravitational coupling 
constant in this dual picture.

In this (dS with boundary)/dS correspondence,
we have mapped the $(d+1)$-dimensional system into $d$-dimensional one.
Since the geometry on the domain wall is again dS, 
we can utilize the dS/CFT correspondence to
obtain a $(d-1)$-dimensional CFT at the future 
infinity~\cite{Strominger:2001pn,Strominger:2001gp}
 (see Fig.~\ref{fig:8}). The possibility of this
second duality was first discussed in~\cite{Garriga:2010fu}.
If we push the system into future timelike infinity,
gravity will decouple from the system. Then, we can describe the whole physics
 in terms of the $(d-1)$-dimensional CFT. 
This seems an interesting issue to study, particularly
in its implications to the measure problem of the multiverse.

\section{Conclusion}
\label{sec:conclusion}

We studied the proposal that an AdS bubble in dS spacetime
can be excized and replaced by a CFT at the boundary. 
In the original picture in which there is an AdS bubble
in the dS spacetime, incident gravitons
 from the dS side can penetrate into the AdS bubble.
We calculated the transmission probability of gravitons through the
bubble wall in arbitrary dimensions. 
In the dual picture, the incident gravitons should be absorbed by 
CFT matter because there is no spacetime beyond the boundary. 
Therefore, the transmission probability in the original picture should be
identical to the absorption probability of gravitons by
CFT matter in the dual picture.

We derived a general formula for the transmission probability,
that is, the absorption probability in the dual picture,
in the decoupling limit of gravity in the AdS bubble,
$a_w/\ell_{\rm AdS}\gg1$, where $a_w$ is the wall radius
and $\ell_{\rm AdS}$ is the AdS curvature radius.
Then we obtained the standard absorption probability in 4-dimensions
in agreement with the result in \cite{Garriga:2010fu},
while we found corrections due to the curvature radius of the 
domain wall in higher dimensions. We argued that
the fact that there were no correction in four dimensions 
is due to the absence of massless (ie, propagating) gravitons
in three dimensions.

We found there is an important difference between
the cases of even and odd spacetime dimensions. 
In odd dimensions, we found thermal poles obeying
Fermi statistics in the amplitude of the absorption probability,
while there were no thermal poles in even dimensions. 
We speculated that this is due to acceleration of the domain 
wall relative to the bulk spacetime, and there exists
a universal factor that cancels thermal poles obeying Bose 
statistics, which happens in even spacetime dimensions,
while the same factor does not cancel thermal poles obeying
Fermi statistics, which is the case in odd spacetime dimensions.
The apparent difference may be understood 
by the breakdown of the Huygens principle.

In order to reveal features of the duality, we studied various situations. 
Here, we summarize those briefly. 

\begin{itemize}
\item {\bf Minkowski disc in AdS spacetime.}\\
In even and odd spacetime dimensions, the absorption probability is 
expressed by Eqs.~(\ref{evenMdisc1}) and (\ref{oddMdisc1}). A correction
caused by the curvature radius of the domain wall, $\lambda^2_{\rm IR}$, 
comes into the formula for both cases and also a correction due to the 
thermal nature of the wall, $\tanh \pi k$, appears in odd spacetime
dimensions. However, if we take the flat wall limit: $k=a_w\, k_w\to\infty$
with $k_w$ fixed, these corrections disappear and the standard absorption
probability in the Minkowski spacetime is obtained
as seen in Eqs.~(\ref{trans:six}) and (\ref{4dimtrans}).
This result is consistent with Maldacena's proposal.

\item {\bf AdS bubble in Minkowski spacetime.}\\
The absorption probability in even and odd spacetime dimensions are given
by Eqs.~(\ref{evenMdisc2}) and (\ref{oddMdisc2}). The expression in even
spacetime dimensions is the same as in the case of a Minkowski disc in AdS 
spacetime. On the other hand in odd spacetime dimensions, we find
a curious additional correction due to the difference in the form of 
the hypergeometric function from that in the previous Minkowski disc case,
the interpretation of which is not immediately clear.
This correction is included in the effective gravitational coupling
as in Eq.~(\ref{effGMin_odd}). 

In spite of these corrections, 
we end up obtaining the standard absorption probability in the Minkowski
spacetime again in the flat wall limit as seen in Eq.~(\ref{4.29}),
which is consistent with Maldacena's proposal.

\item {\bf Maximal radius bubble in dS universe.}\\
In even spacetime dimensions, other than the same correction due to 
$\lambda_{\rm IR}$ as in the previous cases, 
we obtain the effective gravitational coupling constant, 
which is always greater than the original gravitational constant in 
the dS bulk as in Eqs.~(\ref{max:even}) and (\ref{max:odd}). 
The absorption probability is enhanced due to this effective 
gravitational coupling. We find that this enhancement is 
related to the bound states in the dS bulk as discussed in Appendix C.
However since we have $\ell_{\rm dS}=a_w$ by assumption in the present case,
we are unable to identify if the enhanced effect is due to the 
dS curvature outside the bubble or due to the curvature of the domain wall.
Nevertheless, considering the difference from the previous cases
of Minkowski disc or Minkowski bulk,
this correction can be interpreted as the one stemming from the 
dS curvature outside the bubble.
In spite of these corrections, we again obtain
the standard absorption probability in Minkowski spacetime in 
the flat wall limit as in Eq.~(\ref{transmission4}).

In odd spacetime dimensions, it is difficult to obtain a simple,
analytical expression for the absorption probability.
Nevertheless, the product of the absorption probabilities of two
adjacent odd dimensions can be evaluated as in Eq.~(\ref{transmission5}).
The result gives the standard absorption probability in Minkowski spacetime
in the flat wall limit, being consistent with Maldacena's proposal again.

\item {\bf Generic cases.}\\
In order to see the effect of dS curvature more explicitly,
we derived an approximate formula for the general position of 
the domain wall in dS spacetime,
valid for fairly large $k$, where $k=a_w\,k_w$ with
$k_w$ being the graviton energy on the domain wall.
We also checked the formula numerically.
The formula tells us that the effect of the non-vanishing
dS curvature in the bulk increases the absorption probability 
for large $k$ as seen in Fig.~\ref{fig:4}. 
In the case of small $k$, we find resonant peaks 
in the absorption probability as seen in Figs.~\ref{fig:5} and \ref{fig:6}. 
This resonance feature 
is encoded in the effective gravitational coupling constant 
where localized massive gravitons play an important role.

\item {\bf The (dS with boundary)/dS correspondence}\\
The interpretation of this correspondence is natural in that 
bulk gravitons are interpreted as KK gravitons in lower dimensions.
More precisely, our absorption probability formula gives the decay
probability of KK gravitons through two channels,
one directly decaying to CFT matter and the other
indirectly via localized massive graviton modes. 
Here, we mention that the wall fluctuation mode
may have some relevance to localized massive modes,
and discussed in Appendix~\ref{app:wallfluc}.

\end{itemize}

In conclusion, the calculation of the absorption rate of gravitons
shows that Maldacena's excision proposal is physically reasonable.
However, we also find non-trivial correction factors due to
finiteness of the wall radius and/or the dS curvature. 
We discussed the origin and meaning of these corrections.
Apparently it is interesting to understand more clearly
the role of the localized modes of gravitons in this duality.
It may be also interesting to study the relation 
of our approach to other approaches~\cite{Harlow:2010my, Freivogel:2006xu}.

\acknowledgments
We would like to thank Alex Vilenkin for fruitful discussions and
Jaume Garriga for the initiation of this work and useful comments.
SK was supported in part 
by grant PHY-0855447 from the National Science 
Foundation.
This work was supported in part by the
Grant-in-Aid for  Scientific Research Fund of the Ministry of 
Education, Science and Culture of Japan No.22540274, the Grant-in-Aid
for Scientific Research (A) (No.21244033, No.22244030), the
Grant-in-Aid for  Scientific Research on Innovative Area No.21111006,
JSPS under the Japan-Russia Research Cooperative Program,
the Grant-in-Aid for the Global COE Program 
``The Next Generation of Physics, Spun from Universality and Emergence",
and by Korea Institute for Advanced Study under the KIAS Scholar program.

\appendix
\section{Formulas related to associated Legendre functions}
\label{app:Legendre}

The definition of the associated Legendre function $P_\nu^\mu(u)$
we adopt in this paper is
\begin{eqnarray}
P_\nu^\mu(x)=\frac{1}{\Gamma(1-\mu)}
\left(\frac{x+1}{x-1}\right)^{\mu/2}
F\left(-\nu,\nu+1,1-\mu;\frac{1-x}{2}\right)\quad
\mbox{for}~u>1\,,
\end{eqnarray}
and
\begin{eqnarray}
P_\nu^\mu(x)=\frac{e^{i\pi\mu}}{\Gamma(1-\mu)}
\left(\frac{1+x}{1-x}\right)^{\mu/2}
F\left(-\nu,\nu+1,1-\mu;\frac{1-x}{2}\right)\quad
\mbox{for}~u<1\,,
\end{eqnarray}
where $F(a,b,c:u)$ is the hypergeometric function defined by
\begin{eqnarray}
F(a,b,c;u)=
\frac{\Gamma(c)}{\Gamma(a)\Gamma(b)}
\sum_{n=0}^\infty
\frac{\Gamma(a+n)\Gamma(b+n)}{\Gamma(c+n)}\frac{u^n}{n!}\,.
\label{hyperdef}
\end{eqnarray}
This series is known to be convergent for $|u|<1$ on the
complex $u$-plane.

By definition $F(a,b,c;0)=1$ for any $a$, $b$ and $c$.
Hence the behavior of $P_\nu^\mu(x)$ in the limit $x\to1$
is completely governed by the factor 
$\bigl((x+1)/(x-1)\bigr)^{\mu/2}$ or $\bigl((1+x)/(1-x)\bigr)^{\mu/2}$
in front of the hypergeometric function.
In general, $F(a,b,c;u)$ is singular at $u\to1$. However,
it may have a regular limit if both $\Re{c}>0$ and $\Re{c-a-b}>0$
are satisfied. In this case,
\begin{eqnarray}
F(a,b,c;1)=\frac{\Gamma(c)\Gamma(c-a-b)}{\Gamma(c-a)\Gamma(c-b)}\,,
\end{eqnarray}
and the behavior of $P_\nu^\mu(x)$ in the limit $x\to-1$
is governed by the factor 
$\bigl((x+1)/(x-1)\bigr)^{\mu/2}$ or $\bigl((1+x)/(1-x)\bigr)^{\mu/2}$.

The hypergeometric function can be analytically continued to
the region $|u|>1$ of the complex $u$-plane. However in general
there is a branch cut emanating from $u=1$ toward infinity.
Usual convention is to place a branch cut on the real axis; 
$u\in[1,\infty]$.
The analytically continued function can be again expressed in terms
of the hypergeometric functions. There are various analytic continuation
formulas, but one of the most useful formulas is
\begin{eqnarray}
F(a,b,c;u)&=&
\frac{\Gamma(b-a)\Gamma(c)}{\Gamma(b)\Gamma(c-a)}(-u)^{-a}
F(a,a-c+1,a-b+1; 1/u)
\cr
\cr
&&
+\frac{\Gamma(a-b)\Gamma(c)}{\Gamma(a)\Gamma(c-b)}(-u)^{-b}
F(b,b-c+1,b-a+1; 1/u)
\qquad \Bigl(|u|>1\,,~u\neq\mbox{real}\Bigr)\,.
\end{eqnarray}

As clear from the above definition, the dependence of $P_\nu^\mu(x)$
on the lower index $\nu$ is totally encoded in the hypergeometric
function $F$. Since the hypergeometric function has the
symmetry between the first two arguments, $F(a,b,c:u)=F(b,a,c;u)$,
it follows that $P_\nu^\mu(x)$ has the symmetry,
\begin{eqnarray}
P_\nu^\mu(x)=P_{-\nu-1}^\mu(x)\,.
\end{eqnarray}

Another useful property is that the definition of the
hypergeometric function (\ref{hyperdef}) implies
$F(0,b,c;u)=F(b,0,c;u)=1$ independent of $u$, and
if either of $a$ or $b$ is a negative integer,
the series in (\ref{hyperdef}) terminates at a finite order,
that is, it simplifies to a polynomial of a finite order.
Therefore, apart from this finite polynomial factor,
the behavior of $P_\nu^\mu(x)$ is again governed by
the factor 
$\bigl((x+1)/(x-1)\bigr)^{\mu/2}$ or $\bigl((1+x)/(1-x)\bigr)^{\mu/2}$.

Finally, we recapitulate the derivative formula,
\begin{eqnarray}
(1-x^2)\frac{dP_\nu^\mu(x)}{dx}
&=&(\nu+1)\,x\,P_{\nu}^\mu(x)-(\nu-\mu+1)\,P_{\nu+1}^\mu(x)
\cr
&=&(\nu+\mu)\,P_{\nu-1}^\mu(x)-\nu\,x\,P_\nu^\mu(x)\,,
\end{eqnarray}
and the recurrence relation,
\begin{eqnarray}
(\nu+1-\mu)\,P^{\mu}_{\nu+1}(x)
-(2\nu+1)\,x\,P^{\mu}_{\nu}(x)
+(\nu+\mu)\,P^{\mu}_{\nu-1}(x)=0\,.
\end{eqnarray}

\section{Exact transmission probability formula in four dimensions}
\label{app:4Dexact}
Here we derive an exact formula for the transmission probability
in four dimensions without taking the decoupling limit.

Using the function $F_{\nu,k}(x)$ defined in Eq.~(\ref{defF}),
\begin{eqnarray}
F_{\nu,k}(x)=F\left(-\nu,\nu+1,1-ik; \frac{1-x}{2}\right)\,,
\end{eqnarray}
the original general formula for ${\cal M}$, Eq.~(\ref{Minv}),
can be cast in the form,
\begin{eqnarray}
{\cal M}^{-1}
=-e^{-\pi k}\frac{\nu+ik}{\Gamma^2(1-ik)}
\left(\frac{x_0+1}{x_0-1}\right)^{ik/2}
\left(\frac{1+x_0'}{1-x_0'}\right)^{ik/2}
\Bigl(F_{\nu,k}(x_0)F_{\nu-1,k}(x_0')
+F_{\nu-1,k}(x_0)F_{\nu,k}(x_0')\Bigr)\,,
\end{eqnarray}
where we recall that
\begin{eqnarray}
x_0=\coth z_0=\sqrt{\left(\frac{a_w}{\ell_{\rm AdS}}\right)^2+1}\,,
\quad
 x_0'=\tanh z_0'=\pm\sqrt{1-\left(\frac{a_w}{\ell_{\rm dS}}\right)^2}\,.
\end{eqnarray}

Now we focus on the case of four dimensions, $d=3$.
In this case we have $\nu=(d-1)/2=1$, and hence
$F_{\nu-1,k}(x)$ is trivial, $F_{0,k}(x)=1$,
and $F_{1,k}(x)$ simplifies to
\begin{eqnarray}
F_{1,k}(x)=F\left(-1,2,1-ik;\frac{1-x}{2}\right)
=1-\frac{2}{1-ik}\frac{1-x}{2}\,.
\end{eqnarray}
Thus we find
\begin{eqnarray}
{\cal M}^{-1}_{d=3}
&=&-e^{-\pi k}\frac{1+ik}{\Gamma^2(1-ik)}
\left(\frac{x_0+1}{x_0-1}\right)^{ik/2}
\left(\frac{1+x_0'}{1-x_0'}\right)^{ik/2}
\bigl(F_{1,k}(x_0)+F_{1,k}(x_0')\bigr)
\cr
&=&-e^{-\pi k}\frac{1+ik}{\Gamma^2(1-ik)}
\left(\frac{x_0+1}{x_0-1}\right)^{ik/2}
\left(\frac{1+x_0'}{1-x_0'}\right)^{ik/2}
\frac{x_0+x_0'-2ik}{1-ik}
\,.
\end{eqnarray}
Therefore
\begin{eqnarray}
|{\cal M}_{d=3}|^2
=\frac{\pi^2e^{2\pi k}}{\sinh^2\pi k}\frac{k^2}{(x_0+x_0')^2+4k^2}\,.
\end{eqnarray}
Inserting this into the transmission probability 
formula~(\ref{transmission}), we obtain
\begin{eqnarray}
{\cal T}_{d=3}=\frac{4k^2}{(x_0+x_0')^2+4k^2}\,.
\end{eqnarray}
Note that we have not used any approximation, hence this is an exact result,
valid for any $k$ for any values of $x_0$ and $x_0'$.
If we recall the formula for the tension of the wall $T$, 
Eq.~(\ref{tension}), we have
\begin{eqnarray}
x_0+x_0'=\frac{a_w}{\ell_{\rm AdS}}\frac{T}{T_{\rm cr}} 
=\frac{\kappa^2}{2}a_wT\,.
\end{eqnarray}
Therefore the above may be expressed in the form,
\begin{eqnarray}
{\cal T}_{d=3}=\frac{k^2}{A^2+k^2}\,;
\quad 
A\equiv\frac{\kappa^2}{4}a_wT\,.
\end{eqnarray}
We see this completely agrees with Eq.~(29) of Ref.~\cite{Garriga:2010fu}
in the Minkowski bulk limit, $x_0'\to-1$.

\section{Bound states}
\label{app:BS}

Let us consider graviton bound states of Eq.~(\ref{basic}) in 
the $(d+1)$-dimensional spacetime.
In contrast to the propagating solution given by
Eqs.(\ref{soln:ads}) and (\ref{soln:ds}) whose eigenvalue
satisfies $M^2>\nu^2$, a bound state has an eigenvalue $M^2 \leq\nu^2$
(or $k^2\leq 0$), and the boundary condition is that it should vanish at 
the $z \rightarrow \pm \infty$.
Thus, we have 
\begin{eqnarray}
\phi_{\rm AdS} &=& A P^{-\mu}_{\nu} \left( \coth(z_0-z )\right)\,,
\\
\phi_{\rm dS} &=& B P^{-\mu}_{\nu} \left( \tanh(z+z'_0 )\right)\,,
\end{eqnarray}
where
\begin{eqnarray}
\mu = \sqrt{\nu^2-M^2}\,,\quad
\nu\equiv \frac{d-1}{2}\,.
\end{eqnarray}
The two junction conditions given as by Eqs.~(\ref{smooth1}) and 
(\ref{smooth2}) at $z=0$ are now in the form,
\begin{eqnarray}
&& A P^{-\mu}_{\nu} (x_0 ) = B P^{-\mu}_{\nu} (x_0' ) 
\label{bound:smooth1}\,, \\
&& A \left[ \left( \nu + 1 + \mu  \right)P^{-\mu}_{\nu +1} (x_0 )
- (\nu + 1) x_0 P^{-\mu}_{\nu} (x_0 )
- \nu \left( x_0 + x_0' \right) P^{-\mu}_{\nu} (x_0 )
         \right] 
\nonumber\\
&&
 =   B \left[ - \left( \nu + 1 + \mu  \right)P^{-\mu}_{\nu+1} (x'_0 )
+ (\nu+1) x'_0 P^{-\mu}_{\nu} (x'_0 )
         \right] 
\,,\label{bound:smooth2}
\end{eqnarray}
where $x_0=\coth z_0$ and $x_0'=\tanh z_0'$ as defined in
Eq.~(\ref{x0x0pdef}). 
In order to have a non-trivial solution for $A$ and $B$, 
we require the condition,
\begin{eqnarray}
 \left( \nu + 1 + \mu  \right) \left[
 P^{-\mu}_{\nu} (x_0 ) P^{-\mu}_{\nu+1} (x'_0 )
 +P^{-\mu}_{\nu+1} (x_0 ) P^{-\mu}_{\nu} (x'_0 )
 \right]
 =d \left( x_0 + x_0' \right) P^{-\mu}_{\nu} (x_0 )
 P^{-\mu}_{\nu} (x'_0 )   \,.
\end{eqnarray}
Note that if we compare the above with the propagating case,
we see that a bound state corresponds to
a pure imaginary pole at $k=i\mu$ ($\mu\geq0$)
in the transmission coefficient ${\bm T}$ on the complex $k$-plane.

For $x_0 \gg 1$, which is the decoupling limit of gravity
as in Eq.~(\ref{assumption}), the above condition reduces to
\begin{eqnarray}
 \left( \nu - \mu  \right) P^{-\mu}_{\nu - 1} (x'_0 )
  =0 \,,
\label{bsdecgrav}
\end{eqnarray}
where we have used the same recurrence relation given
below Eq.~(\ref{Minv}). 

There is at least one bound state solution given by $\mu =\nu$, 
namely the zero mode solution, $M^2 =0$, independent of the wall
position $x_0'$. In fact the zero mode solution exists for
any value of $x_0$ as well, and it is simply given
in terms of the scale factor as $\phi_0(z)\propto a^\nu(z)$.

In addition to the zero mode, there may be a massive bound
state $0<M^2\leq\nu^2$ if we have
\begin{eqnarray}
P^{-\mu}_{\nu -1} (x'_0 )
= \frac{e^{-\mu \pi i}}{\Gamma(1+\mu )}
\left( \frac{1-x_0'}{1+x_0'}\right)^{\mu/2}
F\left( - \nu + 1 \ , \nu \ , 1+\mu \ ; \frac{1-x_0'}{2} \right)=0\,, 
\label{boundstate}
\end{eqnarray} 
for a given value of $x_0'$.

In the case of four spacetime dimensions, $d=3$ ($\nu=1$), 
the hypergeometric function in the above equation becomes trivial 
$F(0,~1,~c;~x)=1$, hence it is satisfied only when $x_0'=1$.
In this case, however, $F(-\nu+1,~\nu,~1+\mu;~0)=1$ for any value of $\nu$,
hence the dependence of $P_\nu^{-\mu}$ on $\nu$
disappears and we have $P^{-\mu}_{\nu-1}=P^{-\mu}_{\nu}=0$. 
Then from Eq.~(\ref{bound:smooth1}), the solution becomes trivial 
$A=0$. Thus, there is no solution other than the zero mode 
irrespective of $x_0'$ for $d=3$. However, since there is
no dynamical massless graviton in three spacetime dimensions, 
this zero mode is a mathematical artifact. In fact, if we go back to
the original tensor equation in $d$-dimensions, Eq.~(\ref{TTmode}),
one finds there is no dynamical solution for $Y_{ij}$ when $d=3$.

We now try to find massive bound state solutions in higher dimensions
$d>3$ ($\nu>1$) in the three limiting cases of $x_0'=\pm 1$ and $0$. 

\subsection{Minkowski disc limit}
\label{BSMindisc}

For the Minkowski disc limit $x'_0 =1$, we have 
$P^{-\mu}_{\nu-1}=0$ in Eq.~(\ref{boundstate}) irrespective of 
the value of $\mu$.
However, in complete analogy with the $d=3$ case discussed above,
since the $\nu$ dependence of $P^{-\mu}_{\nu-1}$
disappears and we have $P^{-\mu}_{\nu-1}=P^{-\mu}_{\nu}=0$.
Then Eq.~(\ref{bound:smooth1}) gives
a trivial solution $A=0$.
Therefore there is no bound state solution in the Minkowski disc limit
except for the zero mode,
\begin{eqnarray}
M^2 =0 \,.
\end{eqnarray}

\subsection{Maximal radius bubbles}
\label{BSMax}

If the domain wall moves away from $x_0'=1$, massive bound
state solutions may appear for $d>3$. Since an analytical study
is difficult for general values of $x_0'$, we consider the case of 
a maximal radius bubble $x_0'=0$.

In this case, we can solve Eq.~(\ref{boundstate}) analytically
if we recall the formula (\ref{hgformula}).
Setting $a=\nu-1$ and $b=1+\mu$, we find that
the factor $\Gamma((b-a)/2)$ in the denominator diverges at
$\mu - \nu + 2 = -2n'$ ($n'= 0,1,2, \cdots $).
Combining this with the relation $\mu=\sqrt{\nu^2 -M^2}$, 
we obtain the mass spectrum as
\begin{eqnarray}
   M^2 = \nu^2 - \left( \nu - 2n\right)^2;
\qquad n=1,2,\cdots, \left[\frac{\nu}{2}\right]\,,
\end{eqnarray}
where $[Q]$ is the maximum integer not exceeding $Q$.

The above condition determines the number
of massive bound states of gravitons.
In order for the $n$-th massive bound state to exist,
we need $\nu\geq2n$ or $d \geq 4n+1$.
Therefore, the minimum spacetime dimensions to have a massive mode
is six ($d=5$).  As discussed in Section~\ref{subsec:genericcases}, 
there is one massive bound state ($n=1$) in nine dimensions 
($d=8$ or $m=2$ where $d=2m+4$), and two bound states ($n=1,2$)
in ten dimensions ($d=9$ or $m=3$ where $d=2m+3$).

\subsection{Minkowski bulk limit}
\label{BSMinbulk}

For the Minkowski bulk limit $x_0' \to -1$, we use the formula,
\begin{eqnarray}
P^{-\mu}_{\nu-1} (x) &=&
e^{-\mu \pi i} \left( \frac{1-x}{1+x} \right)^{\mu/2}
\left[ \frac{\Gamma(-\mu)}{\Gamma(-\nu+1)\Gamma(\nu)}
\left( \frac{1+x}{2} \right)^{\mu} 
F\left(\mu+\nu, \mu-\nu +1 , 1+\mu ;\frac{1+x}{2}\right)
\right. \nonumber\\
&& {\hskip 4cm}
\left. + \frac{\Gamma(\mu)}{\Gamma(\mu-\nu +1)\Gamma(\mu+\nu)}
F\left(-\nu +1, \nu  , 1-\mu ;\frac{1+x}{2}\right)
\right] \,.
\end{eqnarray}
We see that $\Gamma(\mu-\nu +1)$ in the second term 
in the square brackets diverge at
$\mu = \nu - 1 -n'$ with $n'= 0,1,2, \cdots $.
Then, we obtain the mass spectrum,
\begin{eqnarray}
 M^2 = \nu^2 - \left( \nu -n \right)^2 \,;
\qquad n=1,2,\cdots, \bigl[\nu\bigr]\,.
\end{eqnarray}
The case when $n=\nu$ for an integer $\nu$ needs some care.
Since $\mu=0$ in this case, the above formula is inapplicable.
However, if we remember that the case $\mu=0$ with
integer $\nu$ is simply a Legendre polynomial $P_n(x)$,
we immediately see that $\mu=0$ is not a bound state
solution because $P_n(-1)=(-1)^n\neq0$.

Hence the number of the bound states is $\bigl[\nu-1/2\bigr]$
instead of $\bigl[\nu\bigr]$. Namely, for the $n$-th
bound state to exist we need $\nu\geq(n+1)$ or $d\geq2n+2$.
Thus, the minimum dimensions to have a massive mode in is five
($d=4$). There are three bound states ($n=1,2,3$)
both in nine dimensions ($d=8$ or $m=2$ where $d=2m+4$) and 
in ten dimensions ($d=9$ or $m=3$ where $d=2m+3$).
We note that this result coincides with that of the pure dS case.

In conclusion, we found that there exists a massless bound state 
independent of the position of the domain wall $x_0'$ in all
dimensions higher than four ($d+1\geq5$), while 
there may exist massive bound states and the number 
of them varies depending on the position of the domain wall 
$x_0'$ and the dimension $d$. 
The number of bound states increases as $x_0'$ 
varies from $1$ to $-1$ and also as $d$ increases. 
We have confirmed this with numerical calculations.
As discussed in Section~\ref{subsec:genericcases},
we found numerically that there appears a resonance peak
in the absorption probability every time a new bound state
appears as $x_0'$ varies from $1$ to $-1$ for a given $d$. 

\section{wall fluctuation mode}
\label{app:wallfluc}

First recall that our $(d+1)$-dimensional metric is given by
\begin{eqnarray}
ds^2 = a^2 (z) \left[ dz^2 +  ds_{\rm dS}^2 \right]\,,
\end{eqnarray}
where 
\begin{eqnarray}
ds_{\rm dS}^2=g_{ij}dx^idx^j=-d\tau^2+\cosh^2\tau\,d\Omega_{(d-1)}^2\,,
\end{eqnarray}
 is a dS spacetime with the Ricci tensor given by
\begin{eqnarray}
R_{ij}=(d-1)g_{ij}\,.
\label{dsRicci}
\end{eqnarray}

On this background we place a domain wall at $z=z_w$, and consider
fluctuations of the position of the wall, $z=z_w+\psi(\tau,x^i)$.
Thus $\psi(x^i)$ will behave as a scalar field in $d$ dimensions.
Let the mass of $\psi$ be $M_s$. The field equation is
\begin{eqnarray}
\left[\Box -M_s^2\right]\psi=0\,.
\end{eqnarray}
Now consider a traceless tensor $Q_{ij}$ constructed from $\psi$,
\begin{eqnarray}
Q_{ij}=\left[\nabla_i\nabla_j-\frac{g_{ij}}{d}\Box\right]\psi\,.
\end{eqnarray}
Then
\begin{eqnarray}
\nabla^iQ_{ij}
&=&\nabla^i\left[\nabla_j\nabla_i-\frac{g_{ij}}{d}\Box\right]\psi
=\left[(\nabla_i\nabla_j-\nabla_j\nabla_i)\nabla^i
+\nabla_j\Box-\frac{1}{d}\nabla_j\Box\right]\psi
\cr
&=&\left[R_{ij}\nabla^i+M_s^2\frac{d-1}{d}\nabla_j\right]\psi
=(d-1)\left[1+\frac{M_s^2}{d}\right]\psi\,.
\end{eqnarray}
Hence $Q_{ij}$ becomes transverse-traceless if $M_s^2=-d$.
This tachyonic mode is known to be the wall fluctuation mode
(at least in the limit of Minkowski bulk).

If we calculate $\Box Q_{ij}$ using Eq.~(\ref{dsRicci}),
 we find
\begin{eqnarray}
\Box Q_{ij}=(M_s^2+2d)Q_{ij}\,.
\end{eqnarray}
Comparing this with Eq.~(\ref{TTmode}), we see that
the scalar mass-squared $M_s^2$ would correspond to the tensor
mass-squared $M^2$ as
\begin{eqnarray}
M^2+2=M_s^2+2d\quad
\leftrightarrow\quad
M^2=M_s^2+2d-2\,.
\end{eqnarray}
For the tachyonic mass $M_s^2=-d$, this gives
\begin{eqnarray}
M^2=-d+2d-2=d-2\,.
\end{eqnarray}
Thus the wall fluctuation mode would correspond to the mode $M^2=d-2$.
In terms of $\mu$ with $\mu^2=\nu^2-M^2$, this corresponds to
\begin{eqnarray}
\mu=\sqrt{\nu^2-M^2}=\sqrt{\nu^2-(d-2)}=\frac{d-3}{2}=\nu-1\,.
\end{eqnarray}
The first massive localized mode in the Minkowski bulk limit
obtained in Appendix~\ref{BSMinbulk} is nothing but this 
wall fluctuation mode.
Curiously, this mode disappears in the other two cases.


\begin{thebibliography}{99}

\bibitem{Craps:2010bg}
  B.~Craps,
  arXiv:1001.4367 [hep-th].
  
\bibitem{Maldacena:1997re}
  J.~M.~Maldacena,
  Adv.\ Theor.\ Math.\ Phys.\  {\bf 2}, 231 (1998)
  [Int.\ J.\ Theor.\ Phys.\  {\bf 38}, 1113 (1999)]
  [arXiv:hep-th/9711200].

\bibitem{Gubser:1998bc}
  S.~S.~Gubser, I.~R.~Klebanov and A.~M.~Polyakov,
  Phys.\ Lett.\  B {\bf 428}, 105 (1998)
  [arXiv:hep-th/9802109].

\bibitem{Witten:1998qj}
  E.~Witten,
  Adv.\ Theor.\ Math.\ Phys.\  {\bf 2}, 253 (1998)
  [arXiv:hep-th/9802150].
  
\bibitem{Hertog:2005hu}
  T.~Hertog and G.~T.~Horowitz,
  JHEP {\bf 0504}, 005 (2005)
  [arXiv:hep-th/0503071].

\bibitem{Hertog:2004rz}
  T.~Hertog and G.~T.~Horowitz,
  JHEP {\bf 0407}, 073 (2004)
  [arXiv:hep-th/0406134].

\bibitem{Craps:2007ch}
  B.~Craps, T.~Hertog and N.~Turok,
  arXiv:0712.4180 [hep-th].
  
\bibitem{Craps:2009qc}
  B.~Craps, T.~Hertog and N.~Turok,
  Phys.\ Rev.\  D {\bf 80}, 086007 (2009)
  [arXiv:0905.0709 [hep-th]].

\bibitem{Pretorius:2005gq}
  F.~Pretorius,
  Phys.\ Rev.\ Lett.\  {\bf 95}, 121101 (2005)
  [arXiv:gr-qc/0507014].

\bibitem{Horowitz:2005vp}
  G.~T.~Horowitz,
  JHEP {\bf 0508}, 091 (2005)
  [arXiv:hep-th/0506166].
  
\bibitem{Murata:2007jh}
  K.~Murata, J.~Soda and S.~Kanno,
  Phys.\ Rev.\  D {\bf 75}, 104017 (2007)
  [arXiv:gr-qc/0703029].
          
\bibitem{Maldacena:2010un}
  J.~Maldacena,
  arXiv:1012.0274 [hep-th].
    
\bibitem{Garriga:2010fu}
  J.~Garriga,
  arXiv:1012.5996 [hep-th].

\bibitem{Susskind:2003kw}
  L.~Susskind,
  In *Carr, Bernard (ed.): Universe or multiverse?* 247-266.
  [hep-th/0302219].


\bibitem{Freivogel:2006xu}
  B.~Freivogel, Y.~Sekino, L.~Susskind and C.~P.~Yeh,
  Phys.\ Rev.\  D {\bf 74}, 086003 (2006)
  [arXiv:hep-th/0606204].
            

\bibitem{Garriga:2008ks}
  J.~Garriga and A.~Vilenkin,
  JCAP {\bf 0901}, 021 (2009)
  [arXiv:0809.4257 [hep-th]].
    
\bibitem{Coleman:1980aw}
  S.~R.~Coleman and F.~De Luccia,
  Phys.\ Rev.\  D {\bf 21}, 3305 (1980).


\bibitem{Sato:1981bf}
  K.~Sato, M.~Sasaki, H.~Kodama, K.~-i.~Maeda,
  Prog.\ Theor.\ Phys.\  {\bf 65}, 1443 (1981).
\\
  K.~-i.~Maeda, K.~Sato, M.~Sasaki, H.~Kodama,
  Phys.\ Lett.\  {\bf B108}, 98 (1982).
\\
  K.~Sato, H.~Kodama, M.~Sasaki, K.~-i.~Maeda,
  Phys.\ Lett.\  {\bf B108}, 103 (1982).


\bibitem{Koyama:2001rf}
  K.~Koyama and J.~Soda,
  JHEP {\bf 0105}, 027 (2001)
  [arXiv:hep-th/0101164].
  
\bibitem{Li:2011bt}
  M.~Li and Y.~Pang,
  arXiv:1105.0038 [hep-th].

\bibitem{Ooguri:1985nv}
  H.~Ooguri,
  Phys.\ Rev.\  {\bf D33}, 3573 (1986).

\bibitem{Karch:2003em}
 A.~Karch,
 JHEP {\bf 0307}, 050 (2003)
 [arXiv:hep-th/0305192].

\bibitem{Alishahiha:2004md}
  M.~Alishahiha, A.~Karch, E.~Silverstein and D.~Tong,
  AIP Conf.\ Proc.\  {\bf 743}, 393 (2005)
  [arXiv:hep-th/0407125].
  
\bibitem{Alishahiha:2005dj}
  M.~Alishahiha, A.~Karch and E.~Silverstein,
  JHEP {\bf 0506}, 028 (2005)
  [arXiv:hep-th/0504056].

\bibitem{Strominger:2001pn}
  A.~Strominger,
  JHEP {\bf 0110}, 034 (2001)
  [arXiv:hep-th/0106113].

\bibitem{Strominger:2001gp}
  A.~Strominger,
  JHEP {\bf 0111}, 049 (2001)
  [arXiv:hep-th/0110087].
  
\bibitem{Harlow:2010my}
  D.~Harlow and L.~Susskind,
  arXiv:1012.5302 [hep-th].

\end{thebibliography}
\end{document}